\documentclass{article}
\usepackage{amssymb,latexsym,amsmath,amsthm,graphicx,subfigure}

\advance \topmargin by -\headheight
\advance \topmargin by -\headsep
\evensidemargin \oddsidemargin
\newcommand\Jac{\operatorname{Jac}}

\newtheorem{theorem}{Theorem}

\newtheorem{proposition}[theorem]{Proposition}

\begin{document}
\title{Separation of variables and explicit theta-function solution of the classical Steklov--Lyapunov systems:
A geometric and algebraic geometric background.
\footnote{AMS Subject Classification 37J35, 70E40, 70H06, 70G55, 14H40, 14H42}}
\author{Yuri Fedorov \\
e-mail: Yuri.Fedorov@upc.edu \\
 Department de Matem\`atica Aplicada I, \\
Universitat Politecnica de Catalunya, \\
Barcelona, E-08028 Spain \\
and \\
Inna Basak \\
 Department de Matem\`atica Aplicada I, \\
Universitat Politecnica de Catalunya, \\
Barcelona, E-08028 Spain \\
e-mail: Inna.Basak@upc.edu}
\date{}
\maketitle

\begin{abstract} The paper revises the explicit integration of the classical Steklov--Lyapunov systems via separation of variables,
which was first made by F. K\"otter in 1900, but was not well understood until recently.
We give a geometric interpretation of the separating variables and then,
applying the Weierstrass hyperelliptic root functions, obtain explicit theta-function solution to the problem.
We also analyze the structure of its poles on the corresponding Abelian variety. This enables us to obtain
a solution for an alternative set of phase variables of the systems that has a specific compact form.
\end{abstract}

\section{Introduction}
The motion of a rigid body in the ideal incompressible fluid is described by the classical Kirchhoff equations
$$
\dot K =K \times \frac{\partial H}{\partial K} + p\times \frac{\partial H}{\partial p}, \quad
\dot p = p\times \frac{\partial H}{\partial p},
$$
where $K,p \in{\mathbb R^3}$ are the vectors of the impulsive momentum and the impulsive force, and $H=H(K,p)$
is the Hamiltonian, which is quadratic in $K,p$.
Note that this system always possesses two trivial integrals (Casimir functions of the coalgebra $e^*(3)$)
$\langle K, p\rangle, \langle p, p\rangle$ and the Hamiltonian itself is also a first integral.

Steklov \cite{Steklov} noticed that the classical Kirchhoff equations are integrable under certain conditions i.e., when the Hamiltonian has the form
\begin{equation}\label{1}
H_{1}= \frac{1}{2}  \sum^{3}_{\alpha =1}\Bigl(
b_{\alpha}K_{\alpha}^{2}
+2\nu b_{\beta}b_{\gamma}K_{\alpha}p_{\alpha}
+\nu^{2}b_{\alpha}(b_{\beta}-b_{\gamma})^{2}p^{2}_{\alpha}\Bigr), \qquad (\alpha ,\beta ,\gamma)=(1,2,3)\, ,
\end{equation}
$b_{1},b_{2},b_{3}$ and $\nu$ being arbitrary parameters. Under the Steklov condition, the equations possess
fourth additional integral
\begin{equation}\label{2}
H_{2}= \frac{1}{2}\sum^{3}_{\alpha =1}\Bigl(
K^{2}_{\alpha}-2\nu b_{\alpha}K_{\alpha}p_{\alpha}
+\nu^{2}(b_{\beta}-b_{\gamma})^{2}p^{2}_{\alpha}\Bigr)\, .
\end{equation}
Later Lyapunov \cite{Lyap2} discovered an integrable case of the Kirchhoff
equations whose Hamiltonian was a linear combination of the additional integral \eqref{2} and the two trivial integrals. Thus, the Steklov and Lyapunov integrable systems actually define different trajectories on the same invariant manifolds, two-dimensional tori. This fact was first noticed in \cite{Kolos}.


In the sequel, without loss of generality, we assume $\nu =1$ (this
can always be made by an appropriate rescaling $p\,\rightarrow\,  p/\nu)$.

The Kirchhoff equations with the Hamiltonians
(\ref{1}), (\ref{2}) were first solved explicitly by K\"otter \cite{Kot2}, who used the change of variables $(K,p)\, \rightarrow \, (z,p)$:
\begin{equation}
2z_{\alpha}=K_{\alpha}-(b_{\beta}+b_{\gamma})p_{\alpha}\,  , \qquad
\alpha = 1,2,3\, ,\quad
(\alpha ,\beta ,\gamma)=(1,2,3),
\label{substit}
\end{equation}
which transforms the Steklov--Lyapunov systems to the form
\begin{equation}
\dot z=z\times Bz-Bp\times Bz\, ,\quad \dot p=p\times Bz \, , \qquad B={\rm diag}\,  (b_{1},b_{2},b_{3})
\label{11}
\end{equation}
and, respectively,
\begin{equation}\label{22}
\dot z=p\times Bz\, ,\quad \dot p=p\times (z-Bp)\,  .
\end{equation}

K\"otter implicitly showed that the above systems admit the following Lax representation with $3\times 3$ skew-symmetric matrices
and a spectral parameter
\begin{equation}
\begin{aligned}
\dot L(s)&=[\, L(s),A(s)\, ]\, ,          \qquad
L(s),A(s)\in so(3),\quad s\in \mathbb{C}\, ,       \\
L(s)_{\alpha \beta}
&=\varepsilon_{\alpha \beta \gamma}\Bigl(
\sqrt{s-b_\gamma}\, (z_\gamma + sp_\gamma)\Bigr)\, ,\label{rad1}
\end{aligned}
\end{equation}
where $\varepsilon_{\alpha \beta \gamma}$ is the Levi-Civita
tensor. Equations (\ref{11}) and (\ref{22}) are generated by the operators
\begin{equation}\label{rad2}
A(s)_{\alpha \beta}
= \frac{\varepsilon_{\alpha \beta \gamma}  }{s}
\sqrt{(s-b_\alpha )(s-b_\beta)}\,b_{\gamma}z_{\gamma}\,  , \quad \mbox{resp.} \quad
A(s)_{\alpha \beta}=\varepsilon_{\alpha \beta \gamma}
\sqrt{(s-b_\alpha )(s-b_\beta )}\,  p_{\gamma}.
\end{equation}
The radicals in (\ref{rad1})--(\ref{rad2}) are
single-valued functions on the elliptic curve $\widehat{\cal E}$, the 4-sheeted unramified covering of the plane curve
${\cal E}=\{w^{2}=(s-b_{1})(s-b_{2})(s-b_{3})\}$. For this reason, the Lax representation has an elliptic spectral parameter.

Writing out the characteristic equation for $L(s)$, we arrive at the
 following family of quadratic integrals
\begin{equation}\label{polin}
{\cal F} (s)=\sum^{3}_{\gamma =1} (s-b_\gamma)(z_\gamma +sp_\gamma)^2
\equiv J_{1}s^{3}+J_{2} s^{2}+2sH_{2}-2H_{1} \,  ,
\end{equation}
 where
\begin{equation} \label{ints}
H_{1} =\frac{1}{2} \langle z,Bz \rangle \, ,   \quad
H_{2} =\frac{1}{2} \langle z,z \rangle - \langle Bz,p \rangle \, , \quad J_{2}=2  \langle z,p \rangle- \langle  Bp,p \rangle\, , \quad
J_{1}=  \langle p,p \rangle \, .
\end{equation}

It is seen that under the K\"otter substitution (\ref{substit}) the functions
$J_{1}, J_{2}$ transform into invariants of the coalgebra $e^{*}(3)$, whereas the integrals
$H_{1}(z,p)$, $H_{2}(z,p)$ (up to a linear combination of the invariants) become the Hamiltonians (\ref{1}),(\ref{2}).

An analog of the Lax pair \eqref{rad1} was later rediscovered in \cite{Bob} and was used to obtain theta-function solution of the systems by using the method of Baker--Akhieser functions (see \cite{BE}).
However, the resulting formulas appeared to be quite tedious, and it was not evident how to compare or identify them
with the theta-function solution of K\"otter.

Note that the latter was found in the classical manner, i.e., by a separation of variables and reduction of
the equations of motion to quadratures, which have the form of the Abel--Jacobi map associated to a genus 2 hyperelliptic curve. The phase variables of the Kirchhoff equations have been expressed in terms of the separating variables in a quite symmetric but complicated way. Until recently, various attempts to check these expressions, as well as the reduction to quadratures made by K\"otter, even using packages of modern computer algebra, were not successful.   This even made some specialists to believe that the results of \cite{Kot2} are not reliable hence useless.

One of the first step in verification of K\"otters' calculations was made in \cite{Bols_Fed_2}, where the
Steklov--Lyapunov systems on $e^*(3)$, as well as their higher-dimensional generalizations, have been considered as Poisson reductions of certain Hamiltonian systems in a bigger phase space.
The latter systems were shown to possess $2\times 2$ matrix Lax representations in a generalized Gaudin form with a rational spectral parameter. This fact easily allowed to find separating variables, which coincided with those suggested by K\"otter, and, as a byproduct, prove their commutativity with respect to the Lie-Poisson bracket on $e^*(3)$. A similar approach to the separation of variables was made in \cite{Tsiganov2}.

The main aim of the present paper is to reconstruct the rest of the results of the paper
\cite{Kot2}\footnote{Note that apart from the solutions of the Kirchhoff equations, K\"otter also provided (although in an extremely brief form) the theta-solutions describing the motion of the group $E(3)$, that is, the components of the
rotation matrix of the body and the trajectory of its center in space. We could not reconstruct these solutions.}.

For our purposes we shall also use another set of phase variables which depend linearly on $z,p$. Namely,
putting in \eqref{polin} successively $s=b_{1}$, $s=b_{2}$, $s=b_{3}$ we obtain three independent
quadratic integrals defining rank 3 quadrics in $\mathbb{P}^{6}$:
\begin{equation} \label{rank3}
\begin{aligned}
(b_{1}-b_{2})(z_{2}+b_{1}p_{2})^{2}+(b_{1}-b_{3})(z_{3}+b_{1}p_{3})^{2}
&={\cal F} (b_{1})\, ,                                                      \\
(b_{2}-b_{1})(z_{1}+b_{2}p_{1})^{2}+(b_{2}-b_{3})(z_{3}+b_{2}p_{3})^{2}
&={\cal F} (b_{2})\, ,                                                      \\
(b_{3}-b_{1})(z_{1}+b_{3}p_{1})^{2}+(b_{3}-b_{2})(z_{2}+b_{3}p_{2})^{2}
&={\cal F} (b_{3})\, .
\end{aligned}
\end{equation}
Then it is natural to introduce new variables
\begin{equation} \label{vis}
\begin{aligned}
v_{1} &=\sqrt{(b_2 -b_3)(b_1 -b_2)}\, (z_{2}+b_{1}p_{2})\, ,              \\
v_{2}
&=\sqrt{(b_2 -b_3)(b_3 -b_1)}\, (z_{3}+b_{1}p_{3})\, ,              \\
v_{3}
&=\sqrt{(b_3 -b_1)(b_1 -b_2)}\, (z_{1}+b_{2}p_{1})\, ,              \\
v_{4}
&=\sqrt{(b_2 -b_3)(b_3 -b_1)}\, (z_{3}+b_{2}p_{3})\, ,              \\
v_{5}
&=\sqrt{(b_3 -b_1)(b_1 -b_2)}\, (z_{1}+b_{3}p_{1})\, ,              \\
v_{6}
&=\sqrt{(b_2 -b_3)(b_1 -b_2)}\, (z_{2}+b_{3}p_{2})\,  ,
\end{aligned}
\end{equation}
which, in particular, imply
\begin{gather*}
p_1= \frac{v_3-v_5}{\sqrt{\cal S} \sqrt{b_2-b_3}}, \quad
p_2= \frac{v_1-v_6}{\sqrt{\cal S} \sqrt{b_3-b_1}}, \quad
p_3= \frac{v_2-v_4}{\sqrt{\cal S} \sqrt{b_1-b_2}}, \\
{\cal S}= (b_{1}-b_{2})(b_{2}-b_{3})(b_{3}-b_{1}) .
\end{gather*}
Then the integrals \eqref{rank3} and $(p,p)=J_1$ take the following compact form
\begin{gather}\label{ints_v}
\begin{aligned}
v^{2}_{1}-v^{2}_{2}
&=\psi (b_{1})\, /\, (b_{2}-b_{3})\, ,     \\
v^{2}_{3}-v^{2}_{4}
&=\psi(b_{2})\, /\, (b_{3}-b_{1})\, ,      \\
v^{2}_{5}-v^{2}_{6}
&=\psi (b_{3})\, /\, (b_{1}-b_{2})\, ,
\end{aligned}  \\
{(v_{3}-v_{5})^{2}\over b_{2}-b_{3}} + {(v_{1}-v_{6})^{2}\over b_{3}-b_{1}}
+ {(v_{2}-v_{4})^{2}\over b_{1}-b_{2}} =J_{1} (b_{1}-b_{2})(b_{2}-b_{3})(b_{3}-b_{1})\,  . \notag
\end{gather}

The Steklov--Lyapunov systems written in terms of $v_1,\dots,v_6$, as well as the integrals \eqref{ints_v}, are quite similar to
 those describing the reduction of the integrable geodesic flow on the group $SO(4)$ with the diagonal metric II to the algebra $so(4)$, which was considered
in details in \cite{AvM3, AvM1}. In fact, as was shown by several authors (see e.g., \cite{Bob}), there is
a linear isomorphism connecting the above systems\footnote{On the other hand, one of the Steklov--Lyapunov systems on $e^*(3)$ can also be
regarded as a limit of the system on $so(4)$.}. We shall use this property and the results of \cite{AvM1} to
obtain theta function expressions for the sums and differences of $v_i$, which have an especially simple form.




\section{Separation of variables by F. K\"otter.}
The explicit solution of the Steklov--Lyapunov systems in the generic case was given by K\"otter in the brief communication \cite{Kot2}, where he presented the following scheme.

Let us fix the constants of motion in \eqref{ints}, then the invariant polynomial (\ref{polin}) can be written as \\
\begin{equation} \label{qpolin}
{\cal F}(s)=c_{0}(s-c_{1})(s-c_{2})(s-c_{3}), \qquad c_{0}, c_1, c_2, c_{3}=\mbox{const}.
\end{equation}
Assume, without loss of generality, that $b_{1}<b_{2}<b_{3}$. Then one can show that for real $z,p$ there are two possibilities:

\begin{enumerate}
\item[1)] $c_{1},c_{2},c_{3}$ are all real, then $b_{1} \le c_{1} \le c_{2}
\le c_{3} \le b_{3}$;

\item[2)] $c_1$ is real and $c_2, c_3$ are complex conjugated, then $b_1 \le
c_1 \le b_3$ and either $\rho=\Re c_2=\Re c_3 < b_1$ or $\rho > b_3$.
\end{enumerate}

Next, when no one of $c_\alpha$ coincides with $b_1,b_2,b_3$, the level
variety of the four first integrals of the problem (given by the
coefficients at $s^{3},s^{2},s,s^{0}$) is a union of
two-dimensional tori in ${\mathbb R}^6=(z,p)$.
We restrict ourselves to this generic situation, excluding the other cases, which correspond to periodic or asymptotic motions of the body.

Let $\lambda_{1},\lambda_{2}$ be the roots of the equation
\begin{equation}
f(\lambda)=\sum^{3}_{i=1} {\frac{(z_{j}p_{k}-z_{k}p_{j})^{2}}{\lambda -b_{i}}} =0 \, , \qquad (i,j,k)=(1,2,3) \, ,  \label{8.6}
\end{equation}
where, when all $c_\alpha$ are real,
\begin{equation}
\lambda_{1}\in [ b_{1},c_{1}]\, ,\quad \lambda_{2}\in [ c_{3},b_{3} ]\, .
\label{8.7}
\end{equation}
Then for fixed $c_{0},c_{1},c_{2},c_{3}$
the variables $z,p$ can be expressed in terms of $\lambda_{1},\lambda_{2}$ in such a way that for any $s\in \mathbb{C}$ the following relation holds (see formula (7) in \cite{Kot2})
\begin{equation}
z_{i}+sp_{i}
=\sqrt{c_0} \frac{ x_i \sum\limits_{\alpha=1}^3 (s-c_\alpha) \frac{\sqrt{
-(\lambda_1-c_\alpha)(\lambda_2-c_\alpha)}} {(c_\alpha -c_\beta)(c_\alpha
-c_\gamma)} \biggl( \frac{\sqrt{\Phi (\lambda_1)\psi (\lambda_2)}} {%
(\lambda_1-b_i)(\lambda_2-c_\alpha)} -\frac{\sqrt{\Phi (\lambda_2)\psi
(\lambda_1)}} {(\lambda_2-b_i)(\lambda_1-c_\alpha)} \biggr)} {%
(\lambda_1-\lambda_2) \sum\limits_{\alpha=1}^3 \frac{\sqrt{%
-(\lambda_1-c_\alpha)(\lambda_2-c_\alpha)}} {(c_\alpha -c_\beta)(c_\alpha
-c_\gamma)} } \, ,  \label{k8.8}
\end{equation}
where
\begin{gather}\label{fipsi}
\Phi (\lambda)=(\lambda -b_{1})(\lambda -b_{2})(\lambda -b_{3})\, , \quad
\psi (\lambda)=(\lambda -c_{1})(\lambda -c_{2})(\lambda -c_{3})\, , \\
{x}_{i}= \frac{\sqrt{(\lambda_1-b_i)(\lambda_2-b_i)}} {\sqrt{%
(b_i-b_j)(b_i-b_k)}} \, \label{normx}, \\
(i,j,k) =(1,2,3)\, , \quad (\alpha,\beta,\gamma)=(1,2,3) \, \nonumber .
\end{gather}
Setting in the above expression $s\to \infty$ and $s=0$, one obtains the corresponding formulas for $p_i, z_i$.

Note that for real $z_i, p_i$, in the case (1) (all $c_\alpha$ are real), in view of the condition (\ref{8.7}) all the expressions under the radicals in (\ref{k8.8}) are non-negative. In the rest of the cases the roots can be complex. For any $\alpha=1,2,3$, the branches of $\sqrt{-(\lambda_1-c_
\alpha)(\lambda_2-c_\alpha)}$ in the numerator and the denominator of (\ref{k8.8}) must be the same.

Next, the evolution of $\lambda_{1},\lambda_{2}$ is described by the quadratures
\begin{gather}
\begin{aligned} \frac{d\lambda_1 }{\sqrt{R(\lambda_1)} } +
\frac{d\lambda_2}{\sqrt{R(\lambda_2)}} &= \delta_1\, dt  \, , \\
\frac{\lambda_1\, d\lambda_1}{\sqrt{R(\lambda_1)}} + \frac{\lambda_2\,
d\lambda_2}{\sqrt{R(\lambda_2)}} &=\delta_2\,dt  , \label{8.9} \\
\end{aligned} \\
R(\lambda) =-\Phi (\lambda)\psi (\lambda)  \notag
\end{gather}
with certain constants $\delta_{1},\delta_{2}$ depending on the choice of the Hamiltonian only.
In other words, in the variables $\lambda_1, \lambda_2$ the systems separate. 

Note that the paper \cite{Kot2} does not describe explicitly how to find $\delta_{1},\delta_{2}$. They were calculated in \cite{Bols_Fed_2}, \cite{Tsiganov2}.

The above quadratures rewritten in the integral form
\begin{gather}\label{8.99}
\begin{aligned}
\int_{\lambda_0}^{\lambda_1} \frac{d\lambda }{2\sqrt{R(\lambda) } }
+ \int_{\lambda_0}^{\lambda_2} \frac{d\lambda}
{2\sqrt{R(\lambda)}} & = u_1 ,\\
\int_{\lambda_0}^{\lambda_1} \frac{\lambda \, d\lambda}{2\sqrt{R(\lambda)}}
+ \int_{\lambda_0}^{\lambda_1}\frac{\lambda\, d\lambda}{2\sqrt{R(\lambda)}} & = u_2 \, ,
\end{aligned} \\
u_1=\delta_1 t+ u_{10}, \quad u_2=\delta_2 t+ u_{20} , \label{u_s}
\end{gather}
which represent the Abel--Jacobi map associated to
the genus 2 hyperelliptic curve $\mu^2=- \Phi (\lambda)\psi (\lambda)$.
Inverting the map \eqref{8.99} and substituting symmetric functions of $\lambda_1,\lambda_2, \mu_1, \mu_2$ into (\ref{k8.8}), one finally finds $z,p$ as functions of time.
\medskip

Everyone who had read paper \cite{Kot2} might be surprised by how K\"otter managed to invent the intricate substitution $(z,p)\,\rightarrow\,(\lambda_{1},\lambda_{2},c_{0},c_{1},c_{2},c_{3})$ and to represent the
result in the symmetric form (\ref{k8.8}). Unfortunately, the author of \cite{Kot2} 
gave no explanations of his computations. Nevertheless, it is clear that behind the striking formulas there must be a certain geometric idea, which we try to reconstruct in the next section.

\section{A geometric background of K\"otter's solution.}
Let $(x_{1}: x_{2}: x_{3})$ be homogeneous coordinates in ${\mathbb P}^2$ defined up to multiplication by the same non-zero factor.
Consider a line $l$ in ${\mathbb P}^{2}=(x_{1}:x_{2}:x_{3})$ defined by equation
$$
y_{1}x_{1}+y_{2}x_{2}+y_{3}x_{3}=0 .
$$
Following Pl\"ucker (see e.g., \cite{Grif_Harr}), the coefficients $y_{1},y_{2},y_{3}$ can be regarded as homogeneous coordinates
of a point in the dual projective space $\left( {\mathbb P}^{2}\right)^{\ast }$.
Now let $l_{1},l_{2}$ be two intersecting lines in ${\mathbb P}^2$ with the Pl\"ucker coordinates
$( y^{(1)}_1, y^{(1)}_2, y^{(1)}_3)$, $( y^{(2)}_1, y^{(2)}_2, y^{(2)}_3)$.

Then, for any constants $\lambda ,\mu \in {\mathbb C}$ not vanishing simultaneously, the linear combination
$\lambda y _{\alpha}^{(1)}+\mu y_{\alpha}^{(2)}$ are also Pl\"ucker coordinates of a line $l_{\lambda ,\mu }\subset {\mathbb P}^2$.
Hence, we arrive at an important geometric object, {\it a pencil of lines} in ${\mathbb P}^{2}$, i.e.,
a one-parameter family $l_{\lambda ,\mu }$. It is remarkable that all the lines of a pencil
intersect at the same point ${\mathbf P}\in {\mathbb P}^2$ called {\it the focus} of the pencil.

\begin{theorem}\textup{(\cite{Grif_Harr})} \label{T1}
Let $l_{\lambda ,\mu }$ be a pencil of lines in  ${\mathbb P}^2$ defined by the Pl\"ucker coordinates
 $\lambda y _{\alpha}^{(1)}+\mu y_{\alpha}^{(2)}$, $(\lambda:\mu)\in {\mathbb P}$. Then the homogeneous coordinates of the focus are
$$
{\mathbf {P}}=\left(
y_{2}^{(1)}y_{3}^{(2)}-y_{3}^{(1)}y_{2}^{(2)}:y_{1}^{(1)}y_{3}^{(2)}-y_{3}^{(1)}y_{1}^{(2)}:
y_{1}^{(1)}y_{2}^{(2)}-y_{2}^{(1)}y_{1}^{(2)}\right)\, .
$$
\end{theorem}
Next, consider the family of confocal quadrics in ${\mathbb P}^{2}$
\begin{equation} \label{quadric}
Q(s)=\bigg\{ {\frac{x^{2}_{1}}{s-b_{1}}} + {\frac{x^{2}_{2}}{s-b_{2}}} + {\frac{x^{2}_{3}}{s-b_{3}}} = 0\bigg\}
\end{equation}
and a fixed point $P=(X_{1}:X_{2}:X_{3})$.
Then one defines the spheroconical coordinates $\lambda_{1},\lambda_{2}$ of this point (with respect to $Q(s)$) as the roots of the equation
$$
\frac{X^{2}_{1}}{\lambda-b_{1}} + \frac{X^{2}_{2}}{\lambda-b_{2}} + \frac{X^{2}_{3}}{\lambda-b_{3}}=0.
$$

Now, going back to the Steklov--Lyapunov systems, we make the following observation.
\begin{proposition}
\label{focus_3} The separating variables $\lambda_{1}, \lambda_{2}$ defined by formula (\ref{8.6})
are spheroconical coordinates of the focus $\mathbf{P}$ of the pencil of lines in ${\mathbb P}^2$ with the Pl\"ucker
coordinates $z+sp=(z_1+sp_1: z_2+s p_2: z_3+sp_3)$, $s\in {\mathbb P}$  with respect to the family of quadrics (\ref{quadric}).
\end{proposition}

\noindent{\it Proof.} According to Theorem \ref{T1}, the homogeneous coordinates of the focus $\mathbf{P}$ are
\begin{equation}
(z_{2}p_{3}-z_{3}p_{2}:z_{3}p_{1}-z_{1}p_{3}:z_{1}p_{2}-z_{2}p_{1})\, ,
\label{8.10}
\end{equation}
hence, the spheroconical coordinates of $\mathbf{P}$ with respect to the family (\ref{quadric}) are precisely the roots of the equation
\eqref{8.6}, i.e., $\lambda_{1}, \lambda_{2}$. $\square$
\medskip

Note also the following property: for $\alpha=1,2,3$, the line $\ell_{\alpha}$ with the Pl\"ucker coordinates $z+c_{\alpha}p$ is tangent to the quadric
$Q_\alpha= Q(c_{\alpha})$. Indeed, setting in the right hand side of (\ref{polin}) $s=c_{\alpha}$, we obtain
$$
 \sum^{3}_{i=1} (c_\alpha -b_{i})(z_{i}+ c_\alpha p_{i})^{2}=0 \, ,
$$
which represents the condition of tangency of the line $\ell_{\alpha}$ and the quadric $Q_{\alpha}$.

As a result, the following configuration holds: {\it the three lines $\ell_{1},\ell_{2},\ell_{3}$ in $\mathbb P^{2}$ intersect at the same
point $\bf P$ and are tangent to the quadrics $Q_{1},Q_{2},Q_{3}$ respectively}. An example of such a configuration is shown in Fig.~1.

It follows that a solution $z(t)$, $p(t)$ defines a trajectory of the focus $\mathbf{P}$ on $\mathbb{P}^{2}$ or on
$S^{2}=\{ x^{2}_{1}+x^{2}_{2}+x^{2}_{3}=1 \}$, and it natural to suppose that
the Steklov--Lyapunov systems define dynamical systems on the sphere. Indeed, some of these systems were studied in \cite{Tsiganov2}
and were shown to be related to a generalization of the classical Neumann system with an additional quartic potential.

\begin{figure}[ht]
\begin{center}
\includegraphics[width=.7\textwidth]{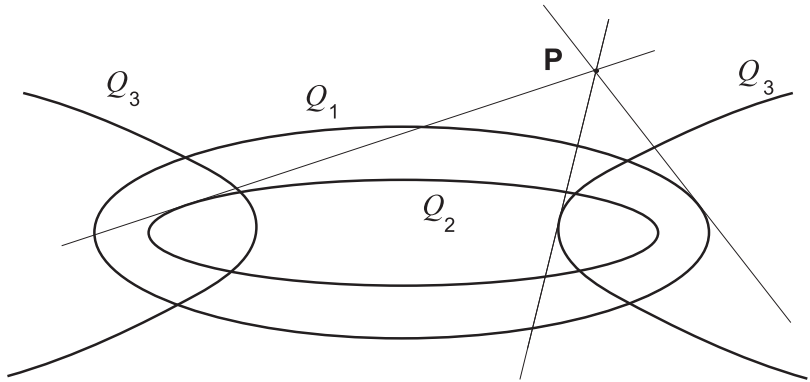}
\end{center}
\caption{A configuration of tangent lines in
${\mathbb R}^{2}=\left(\frac{x_{1}}{x_{3}},\; \frac{x_2}{x_3} \right)$ for the case
$b_1 < c_1 < b_2 <c_2 <c_3< b_3$, when the quadrics $Q_\alpha$ are two ellipses and a hyperbola.}
\label{fig.13}
\end{figure}

\medskip

In the sequel our main goal will be to recover the variables $z$ and $p$ as functions of the spheroconical coordinates of the focus $\mathbf{P}$,
that is, to reconstruct the K\"otter formula (\ref{k8.8}). Obviously, the solution is not unique: due to square roots in \eqref{normx}, each pair
$(\lambda_{1},\lambda_{2}), \lambda_{k}\neq b_{1},b_{2},b_{3}$ gives 4 points on $\mathbb{P}^{2}$, and for each point
$\mathbf{P}$ that does not lie on any of the quadrics $Q(c_{\alpha})$, $2^{3}=8$ different configurations of tangent lines
$\ell_{1},\ell_{2},\ell_{3}$ are possible (Fig.~1 shows just one of them). Thus, under the above generality conditions,
a pair $(\lambda_{1},\lambda_{2})$ gives 32 different tangent configurations.

\paragraph{Reconstruction of $z,p$ in terms of the separating variables.}
Let $(\mathbb {P}^{2})^{*}=(G_{1}:G_{2}:G_{3})$ be the dual space to ${\mathbb P}^2=(x_{1}:x_{2}:x_{3})$, $(G_{i}$
being the Pl\"ucker coordinates of lines in $\mathbb {P}^{2})$. It is convenient to regard $G_{i}$ also as Cartesian
coordinates in the space $(\mathbb{C}^{3})^{*}=(G_{1},G_{2},G_{3})$. The
pencil $\sigma (\mathbf{P})$ of lines in $\mathbb{P}^{2}$ with the focus (\ref{8.10})
is represented by a line in $(\mathbb {P}^{2})^*$ or by plane
$$
\pi=\{(z_{2}p_{3}-z_{3}p_{2}) G_{1}+ (z_{3}p_{1}-z_{1}p_{3}) G_{2}+ (z_{1}p_{2}-z_{2}p_{1}) G_{3}=0\}\subset (\mathbb{C}^{3})^{*}.
$$
Consider the line $\bar\sigma(\mathbf{P})=\{z+sp| s\in {\mathbb R}\}\subset (\mathbb{C} ^{3})^{*}$.
Obviously, $\{z+sp\}\subset \pi $. Now let us use
the condition for the three lines $\ell_{1},\ell_{2},\ell_{3}$ defined by
the points $z+c_{1}p$, $z+c_{2}p$, $z+c_{3}p$ in $(\mathbb{P}^{2})^*$ to be
tangent to the quadrics $Q(c_{1})$, $Q(c_{2})$, $Q(c_{3})$ respectively. Let
${\mathbf V}_{\alpha}=(V_{\alpha 1},V_{\alpha 2},V_{\alpha3})\subset \pi$, $\alpha=1,2,3$ be some vectors in $(\mathbb{C}^{3})^{*}$
representing these points, so that
$\ell_{\alpha}=\{V_{\alpha 1}x_{1}+{V}_{\alpha 2}x_{2}+{V}_{\alpha 3}x_{3}=0\}$.
Then we have 
\begin{equation}
 z+c_{1}p-\mu_{1}{\mathbf V}_{1}=0\, ,\quad z+c_{2}p-\mu_{2}{\mathbf V}_{2}=0\,
,\quad z+c_{3}p-\mu_{3}{\mathbf V}_{3}=0 \kern-2em  \label{8.11}
\end{equation}
for some indefinite factors $\mu_{\alpha}$. This system is equivalent to a
homogeneous system of 9 scalar equations for 9 variables $
z_{\alpha},p_{\alpha},\mu_{\alpha}$, $\alpha =1,2,3$.
Thus the variables can be found up to multiplication by a common factor.
Eliminating $z,p$ from (\ref{8.11}), we obtain the following homogeneous system for $\mu_{1},\mu_{2},\mu_{3}$
\begin{equation*}
(c_{2}-c_{3})V_{\alpha 1}\mu_{1}+(c_{3}-c_{1})V_{\alpha 2}\mu
_{2}+(c_{1}-c_{2})V_{\alpha 3}\mu_{3}=0\, ,\quad \alpha=1,2,3\, ,
\end{equation*}
which has a nontrivial solution, since $\det \Vert V_{\alpha i}\Vert =0$
(the vectors ${\mathbf V}_{\alpha}$  lie in the same hyperplane $\pi)$. It follows,
for example, that
\begin{gather}
\mu_{1} =\mu \varSigma_{1}/(c_{2}-c_{3})\, , \quad \mu_{2} =\mu \varSigma%
_{2}/(c_{3}-c_{1})\, ,\quad \mu_{3} =\mu \varSigma_{3}/(c_{1}-c_{2})\, ,
\label{8.12} \\
\varSigma_{1}=V_{22}V_{33}-V_{32}V_{23}\, , \quad \varSigma_{2}=V_{32}V_{13}-V_{33}V_{12}\, , \quad \varSigma_{3}=V_{12}V_{23}-V_{13}V_{22}\, ,
\label{varsigmas}
\end{gather}
$\mu \neq 0$ being an arbitrary factor. Substituting these expressions into (%
\ref{8.11}) and using the obvious identity
\begin{equation*}
\varSigma_{1}{\mathbf V}_{1}+\varSigma_{2}{\mathbf V}_{2}+\varSigma_{3}{\mathbf V}_{3}=0 \, ,
\end{equation*}
after transformations we find
\begin{align}
p &= {\frac{\mu }{(c_{1}-c_{2})(c_{2}-c_{3})(c_{3}-c_{1})}} (c_{1}\varSigma%
_{1}{\mathbf V}_{1}+c_{2}\varSigma_{2}{\mathbf V}_{2}+c_{3}\varSigma_{3}{\mathbf V}_{3})\, ,  \label{sum_p}
\\
z &= {\frac{\mu }{(c_{1}-c_{2})(c_{2}-c_{3})(c_{3}-c_{1})}} (c_{2}c_{3}%
\varSigma_{1}{\mathbf V}_{1}+ c_{1}c_{3}\varSigma_{2}{\mathbf V}_{2}+ c_{1}c_{2}\varSigma%
_{3}{\mathbf V}_{3})\, .
\end{align}
As a result,
\begin{equation}
\null\kern-4em z+sp = \frac{\mu}{(c_{1}-c_{2})(c_{2}-c_{3})(c_{3}-c_{1})}
\sum_{\alpha=1}^3 (c_{\alpha}s+c_{\beta}c_{\gamma})\varSigma%
_{\alpha}{\mathbf V}_{\alpha} \, .  \label{8.13}
\end{equation}

Now we express the components of ${\mathbf V}_{\alpha}$ in terms of $\lambda_1, \lambda_2$. Up to an arbitrary nonzero
factor, they can be found from the system of equations
\begin{equation}
V_{\alpha 1} x_{1}+V_{\alpha 2} x_{2}+V_{\alpha3} x_{3}=0\, , \quad
\sum^{3}_{i=1}(c_{\alpha}-b_{i})V^{2}_{\alpha i}=0 \, , \qquad \alpha=1,2,3, \label{8.14}
\end{equation}
which represent the conditions that the line $\ell_{\alpha}$ passes through
the focus $\mathbf{P}=(x_{1}: x_{2}: x_{3})$ and touches the quadric $%
Q(c_{\alpha})$.

In the sequel we apply the normalization $x^{2}_{1}+x^{2}_{2}+x^{2}_{3}=1$, which gives rise to expressions  \eqref{normx}.

For $\mathbf{P}\notin Q(c_{\alpha}) $, this system possesses
two different solutions, and for $\mathbf{P}\in Q(c_{\alpha})$ a single one
(the line touches $Q(c_{\alpha})$ at the point $\mathbf{P}$). In the latter
case we can just put
\begin{equation}
V_{\alpha i}= x_{i}\, /\, (c_{\alpha}-b_{i})\, .  \label{8.15}
\end{equation}

Next, it is obvious that under reflection $(x_{1}: x_{2}: x_{3})\,
\rightarrow \, (- x_{1}: x_{2}: x_{3})$,  a solution $(V_{\alpha
1}:V_{\alpha 2}:V_{\alpha 3})$ transforms to $(-V_{\alpha 1}:V_{\alpha
2}:V_{\alpha 3})$ (similarly, for the two other reflections). Let us seek
solutions of equations (\ref{8.14}) in the form of symmetric functions of
the complex coordinates $\lambda_{1},\lambda_{2}$ such that

\begin{enumerate}
\item[1)]  for $\lambda_{1}=c_{\alpha}$ or $\lambda_{2}=c_{\alpha}$ (i.e.,
when $\mathbf{P}\in Q(c_{\alpha})$) there is a unique solution proportional
to (\ref{8.15});

\item[2)]  if $\lambda_{1}$ or $\lambda_{2}$ circles around the point $%
\lambda=c_{\alpha}$ on the complex plane $\lambda$, the two solutions
transform into each other;

\item[3)]  for $\lambda_{1}=b_{i}$ or $\lambda_{2}=b_{i}$ (i.e., when ${x}%
_{i}=0)$, $V_{\alpha i}$ does not vanishes.
\end{enumerate}

Using the Jacobi identities
\begin{equation} \label{jacobi}
\sum\limits_{i=1}^{n}\frac{a^{k}_i}{\prod \left( a_{i}-a_{j}\right) }
=\left\{
\begin{array}{c}
0, \quad k<n-1 \\
1, \quad k=n-1 \\
\sum\limits_{i=1}^{n}a_{i},\quad k=n,
\end{array}
\right.
\end{equation}
one can check that the following expressions satisfy equations (\ref{8.14}) and the above three conditions
\begin{equation}  \label{8.16}
V_{\alpha i} =x_{i}\biggl( \frac{\sqrt{\Phi (\lambda_1)(\lambda_2-c_\alpha )}
} {\lambda_{1}-b_{i}} +\frac{\sqrt{\Phi (\lambda_2)(\lambda_1-c_\alpha)} }{%
\lambda_{2}-b_{i}} \, \biggr)\, , \quad
x_{i} = \frac{ \sqrt{(\lambda_1 -b_{i})(\lambda_2-b_i)}}{ \sqrt{(b_i-b_j)(b_i-b_k)}}.
\end{equation}
Then, using again the identities (\ref{jacobi}), we have
\begin{equation}
\langle {\mathbf V}_{\alpha},{\mathbf V}_{\beta} \rangle \equiv (\lambda_{2}-\lambda_{1}) %
\biggl( \sqrt{(\lambda_2-c_\alpha)(\lambda_2-c_\beta)} - \sqrt{%
(\lambda_1-c_\alpha)(\lambda_1-c_\beta)} \biggr) \, .  \label{8.17}
\end{equation}
and, in particular, $\langle {\mathbf V}_{\alpha},{\mathbf V}_{\alpha}
\rangle=(\lambda_{1}-\lambda_{2})^{2}$ for $\alpha=1,2,3$.

Next, substituting (\ref{8.16}) into (\ref{varsigmas}) and applying the symbolic multiplication rule $\sqrt{ab}\sqrt{ac}=a \sqrt{bc}$, we
find the factors $\varSigma_{\alpha}$ in form
\begin{gather}
\varSigma_{\alpha} = (\lambda_{1}-\lambda_{2}) {x}_{1} \Bigl( \sqrt{-(\lambda_{1}-c_\gamma)(%
\lambda_{2}-c_\beta)} -\sqrt{-(\lambda_1-c_\beta)(\lambda_2-c_\gamma)}
\Bigl)\, ,  \label{8.18} \\
(\alpha ,\beta ,\gamma)=(1,2,3)\, .  \notag
\end{gather}

Further, putting (\ref{8.16}), (\ref{8.18}) into (\ref{8.13}), we obtain
\begin{align}
z_i+sp_i & = \frac{\mu (\lambda_1-\lambda_2) x_1 }{%
(c_{1}-c_{2})(c_{2}-c_{3})(c_{3}-c_{1})} x_i \cdot \sum_{\alpha=1}^3
(c_{\alpha}s +c_\beta c_\gamma)  \notag \\
& \cdot \left [\tfrac {\sqrt {\Phi(\lambda_1 ) \psi(\lambda_2 )} }{%
\lambda_1-b_i} \left (\sqrt{ \tfrac {\lambda_1 -c_\gamma}{ \lambda_2
-c_\gamma} } -\sqrt{ \tfrac {\lambda_1 -c_\beta }{ \lambda_2 -c_\beta} }
\right) + \tfrac {\sqrt {\Phi(\lambda_2 ) \psi(\lambda_1 )} }{\lambda_2-b_i}
\left ( \sqrt{\tfrac {\lambda_2 -c_\gamma}{ \lambda_1 -c_\gamma} } - \sqrt{
\tfrac {\lambda_1 -c_\beta}{ \lambda_2 -c_\beta} } \right) \right ]  \notag
\\
&\equiv \mu (\lambda_{1}-\lambda_{2}) {x}_{1} x_i \sum\limits_{\alpha=1}^3 (s-c_\alpha) \tfrac{%
\sqrt{-(\lambda_1-c_\alpha)(\lambda_2-c_\alpha)}} {(c_\alpha -c_\beta)
(c_\alpha -c_\gamma)} \biggl( \tfrac{\sqrt{\Phi (\lambda_1)\psi (\lambda_2)}%
} {(\lambda_1-b_i)(\lambda_2-c_\alpha)} - \tfrac{\sqrt{\Phi (\lambda_2)\psi
(\lambda_1)}} {(\lambda_2-b_i)(\lambda_1-c_\alpha)} \biggr) \, ,
\label{almost_kotter}
\end{align}
which, up to multiplication by a common factor, coincides with the numerator in K\"otter's formula (\ref{k8.8}).

To determine the factor $\mu $ in (\ref{8.13}) and in (\ref{almost_kotter}), we apply the condition $%
\langle p,p \rangle =c_{0}$ which follows from (\ref{qpolin}). Then, from \eqref{sum_p} we get
\begin{equation}  \label{mu_0}
\frac{c_{0}}{\mu^{2}} = \frac{ |c_{1}\varSigma_{1}{\mathbf V}_{1}+c_{2}\varSigma%
_{2}{\mathbf V}_{2}+c_{3}\varSigma_{3}{\mathbf V}_{3}|^{2}} {
(c_{1}-c_{2})^{2}(c_{2}-c_{3})^{2}(c_{3}-c_{1})^{2} } \, .
\end{equation}
Using the expressions \eqref{8.17}, we obtain
\begin{align*}
\left| \sum_{\alpha=1}^3 c_\alpha \Sigma_\alpha {\mathbf V}_\alpha \right|^2 &
\equiv\sum_{\alpha=1}^3 \left[ c_\alpha^2 \varSigma_{\alpha}^2 \langle
{\mathbf V}_{\alpha},{\mathbf V}_{\alpha}\rangle +2 c_\beta c_\gamma \varSigma_{\beta} \varSigma%
_{\gamma} \langle {\mathbf V}_{\beta},{\mathbf V}_{\gamma}\rangle \right] \\
= (\lambda_1-\lambda_2)^3 x_1^2 & \sum_{\alpha=1}^3 \bigg [ c_\alpha^2
(\lambda_1-\lambda_2) \left( \sqrt{-(\lambda_{1}-c_\gamma)(\lambda_{2}-c_%
\beta)} -\sqrt{-(\lambda_1-c_\beta)(\lambda_2-c_\gamma)} \right)^2 \\
& \qquad +2 c_\beta c_\gamma \left( \sqrt{-(\lambda_{1}-c_\gamma)(%
\lambda_{2}-c_\beta)} -\sqrt{-(\lambda_1-c_\beta)(\lambda_2-c_\gamma)}
\right) \\
& \qquad \cdot \left( \sqrt{-(\lambda_{1}-c_\alpha)(\lambda_{2}-c_\gamma)} -%
\sqrt{-(\lambda_1-c_\gamma)(\lambda_2-c_\alpha)} \right) \\
& \qquad \cdot \left( \sqrt{-(\lambda_{2}-c_\beta)(\lambda_{2}-c_\gamma)} -%
\sqrt{-(\lambda_1-c_\beta)(\lambda_1-c_\gamma)} \right) \bigg ]\, .
\end{align*}
Simplifying the above expression and again using symbolic multiplication of
square roots, one can verify that it is a full square:
\begin{equation*}
\left| \sum_{\alpha=1}^3 c_\alpha \Sigma_\alpha {\mathbf V}_\alpha \right|^2 = x_1^2 (\lambda_1- \lambda_2)^4 \left( \sum^{3}_{\alpha=1} (c_\beta-c_\gamma)
\sqrt{-(\lambda_1-c_\alpha)(\lambda_2-c_\alpha) } \right)^2 .
\end{equation*}
Hence, from \eqref{mu_0} we find
\begin{equation*}
\frac {\sqrt{c_0} }{\mu} =x_1 (\lambda_1- \lambda_2)^2 \sum^{3}_{\alpha=1}
\frac{\sqrt{-(\lambda_1-c_\alpha)(\lambda_2-c_\alpha)}} {(c_\alpha
-c_\beta)(c_\alpha -c_\gamma)} .
\end{equation*}
Combining this with \eqref{almost_kotter}, we finally arrive at (\ref{k8.8}).

Thus, we derived the remarkable K\"otter formula by making use of the
geometric interpretation of the variables $\lambda_{1},\lambda_{2}$. We also
note that the expressions (\ref{k8.8}) are symmetric in $\lambda_{1}, \lambda_{2}$.
\medskip

\paragraph{Remark 1.} As noticed above, a disordered generic pair
$(\lambda_{1},\lambda_{2})$ gives 32 different configurations of tangent lines to the quadrics $Q(c_{1})$, $Q(c_{2})$, $Q(c_{3})$.
Since the common factor $\mu$ in (\ref{8.13}) is defined up to sign flip,
we conclude that, according to the formula (\ref{k8.8}), \emph{to each generic pair}
$(\lambda_{1},\lambda_{2})$ \emph{there correspond} 64 \emph{different points}
$(z,p)$ \emph{on the invariant manifold (a union of 2-dimensional tori) defined by the constants}
$c_{0},c_{1},c_{2},c_{3}$. 
This ambiguity corresponds to different signs of the square roots in the K\"otter formula.
\medskip

In the next section we shall use the expressions (\ref{k8.8}) and the quadratures (\ref{8.99})
to find explicit theta-functional solutions for the Steklov--Lyapunov systems.

\section{Explicit theta-function solution of the Steklov-Lyapunov systems}

In order to give explicit theta-functions solution, we first recall some basic formulas describing inversion of the quadratures (\ref{8.9}).
We shall mainly follow the description given in \cite{Baker1, BE, Dub_NE}.
Consider an even order hyperelliptic Riemann surface of genus $g$ represented in the standard form
$$
\Gamma =\left\{ \mu ^{2}=\left( \lambda -E_{1}\right)\cdots \left( \lambda
-E_{2g+2}\right) \right\}\ \in {\mathbb C}^2(\lambda,\mu).
$$
In the sequel we shall regard $\Gamma$ as its complex compactification obtained by gluing two infinite points $\infty_-, \infty_+$,
where the coordinate $\lambda$ equals infinity.

Consider also differential 1-form (differential) $\omega=\phi(\tau)d\tau$ on $\Gamma$, where $\tau$ is a local parameter at a point $P\in \Gamma$.
A differential $\omega$ is called holomorphic if $\phi(\tau)$ is a holomorphic function for any point $P$.
We choose the canonical basis of cycles ${\mathfrak a}_1, \dots ,{\mathfrak a}_g,{\mathfrak b}_1, \dots ,{\mathfrak b}_g$ on the surface $\Gamma $
such that their intersections are of the form:
$$
{\mathfrak a}_i\circ {\mathfrak a}_j={\mathfrak b}_i\circ {\mathfrak b}_j=0,\quad {\mathfrak a}_i\circ {\mathfrak b}_j=\delta_{ij},\qquad i,j=1,\dots ,g,
$$
where $\gamma_{1}\circ \gamma_{2}$ denotes the intersection index of the cycles
$\gamma_{1},\gamma_{2}$.

\begin{figure}[h,t]
\begin{center}
\includegraphics[height=0.25\textwidth, width=0.75\textwidth]{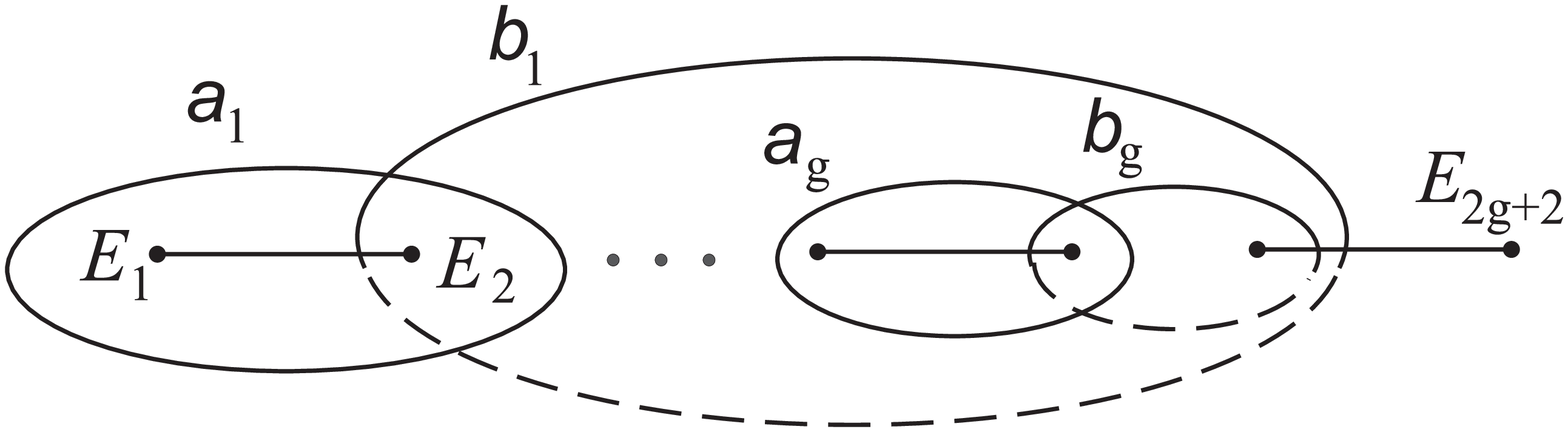}
\end{center} \caption{} \label{fig.2}
\end{figure}

An example of a canonical basis of cycles on $\Gamma$ is shown on Figure 2.
The parts of the cycles on the lower sheet are shown by dashed lines.

Next, let $\bar{\omega}_{1},\dots,\bar{\omega}_{g}$
be the conjugated basis of {\it normalized} holomorphic differentials on $\Gamma$ such that
$$
\oint_{{\mathfrak a}_j}{\bar\omega}_{i}=2\pi \jmath \,\delta_{ij},
\qquad \jmath =\sqrt{-1}.
$$
The $g\times g$ matrix of $b$-periods $B_{ij}=\oint_{{\mathfrak b}_j}\bar\omega_{i}$
is symmetric and has a negative definite real part. Consider the period lattice
$\Lambda^0=\{2\pi \jmath{\mathbb Z}^{g}+B{\mathbb Z}^{g}\}$ of rank $2g$
 in ${\mathbb C}^{g}=(z_1,\dots,z_g)$.
The complex torus Jac$(G)={\mathbb C}^{g}/\Lambda^0$ is called the Jacobi
variety ({\it Jacobian}) of the curve $G$.

Now consider a generic divisor of points
$P_{1}=(\lambda _{1},\mu_{1}),\ldots, P_{g}=(\lambda _{g},\mu_{g})$
on it, and the Abel--Jacobi mapping with a basepoint $P_0$
\begin{gather}
\label{1.16}
\int^{P_{1}}_{P_{0}}\bar{\omega}+\cdots+\int^{P_{g}}_{P_{0}}\bar{\omega}=z, \\
\bar{\omega}=(\bar{\omega}_{1},\dots,\bar{\omega}_{g})^{T}, \quad
z=(z_{1},\ldots,z_{g})^{T}\in {\mathbb C}^{g} .\nonumber
\end{gather}
Under the mapping,  functions on $S^{g}\Gamma$, i.e., symmetric functions of the points
$P_{1},\ldots,P_{g}$ are $2g$-fold periodic functions of the complex variables
$z_{1},\ldots,z_{g}$ with the above period lattice $\Lambda^{0}$ (Abelian functions).

Explicit expressions of such functions can be obtained
by means of theta-functions on the universal covering  ${\mathbb C}^g=(z_{1},\ldots,z_{g})$
of the complex torus.
Recall that customary Riemann's theta-function $\theta(z|B)$ associated with the
Riemann matrix $B$ is defined by the series\footnote{The expression for $\theta(z)$ we use
here is different from that chosen in a series of books on theta-functions by
multiplication of $z$ by a constant factor.}
\begin{gather}
\theta (z|B)= \sum_{M\in {\mathbb Z}^{g}}\exp ( \langle BM,M\rangle +\langle M,z\rangle ),  \label{1.theta-def}  \\
\langle M,z\rangle =\sum^{g}_{i=1}M_{i}z_{i}, \quad
\langle BM,M\rangle =\sum^{g}_{i,j=1}B_{ij}M_{i}M_{j} .\nonumber
\end{gather}
Equation $\theta(z|B)=0$ defines a codimension one
subvariety $\Theta\in \mbox{Jac} (\Gamma)$ (for $g>2$ with singularities) called {\it theta-divisor}.

We shall also use theta-functions with characteristics
$\alpha =(\alpha_{1},\ldots,\alpha_{g})$,
$\beta =(\beta_{1},\ldots,\beta_{g})$, $\alpha_j,\beta_j \in{\mathbb R}$,
which are obtained
from $\theta (z|B)$ by shifting the argument $z$ and multiplying by
an exponent\footnote{Here and below we omit $B$ in the theta-functional notation.}:
$$
\theta\! \left[{ \alpha \atop \beta}\right]\! (z)
\equiv \theta \! \left[{\alpha_{1}\, \cdots \, \alpha_{g} \atop
                       \beta_{1} \, \cdots \, \beta_{g}} \right]\! (z)
=\exp \{\langle B\alpha ,\alpha \rangle /2+\langle z+2\pi  \jmath\beta,\alpha \rangle \}\,
\theta (z+2\pi \jmath\beta+B\alpha) .
$$
Then for a pair of characteristics one has the following useful relations
\begin{equation}\label{*.*}
\theta\! \left[ { \alpha+\alpha' \atop \beta+\beta' } \right ] \!(z)
=\exp \{(B\alpha',\alpha')/2+(z+2\pi \jmath\beta+2\pi \jmath\beta',\alpha')\}\,
\theta \! \left[ {\alpha \atop \beta } \right ] \!
(z+2\pi \jmath\beta'+B\alpha').
\end{equation}
All these functions possess the quadiperiodic property
\begin{eqnarray}
\theta\! \left[ { \alpha \atop \beta } \right ] \! (z+2\pi \jmath K+BM)
=\exp (2\pi \jmath\epsilon)\exp \{-\langle BM,M\rangle /2-\langle M,z\rangle \}
\theta\! \left[ { \alpha \atop \beta } \right ]  \!(z) , \label{1.5} \\
\epsilon =\langle \alpha ,K\rangle -\langle \beta ,M\rangle  , \nonumber
\end{eqnarray}

An important particular case is represented by theta-functions with
half-integer characteristics
$$
\Delta=\begin{pmatrix} \Delta' \\ \Delta'' \end{pmatrix}, \quad
\eta_i=\begin{pmatrix} \eta_i' \\ \eta''_i \end{pmatrix}, \quad \mbox{and} \quad
\eta_{ij}=\eta_{i}+\eta_{j} \quad ({\rm mod} \; {\mathbb Z}^{2g}/{\mathbb Z}^{2g}), \quad
\Delta' ,\Delta'',   \eta_i' , \eta''_i \in \frac 12 {\mathbb Z}^g/{\mathbb Z}^g
$$
such that
\begin{align}
2\pi \jmath \, \eta_i''+B\eta_i' & = \int^{E_i}_{E_{2g+2}}\bar{\omega}
\quad ({\rm mod} \;\Lambda),   \label{eta} \\
 2\pi \jmath  \Delta''+B\Delta' & ={\cal K} \; ({\rm mod}\; \Lambda), \nonumber
\end{align}
${\cal K}\in {\mathbb C}^g$ being the vector of the Riemann constants and $E_{i}$ briefly denotes the branch point $(E_{i},0)$ on $\Gamma$.

The half-integer characteristic $\left[ { \alpha \atop \beta } \right ] \!$ is odd (even) if $\theta\! \left[ { \alpha \atop \beta } \right ] \!(z)$ is odd
(respectively, even).

For the case $g=2$ and for the chosen canonical basis of cycles
${\mathfrak a}_1, {\mathfrak a}_2,{\mathfrak b}_1, {\mathfrak b}_2$ on $\Gamma$
the above characteristics $\Delta ,\eta _{i}$ are
\begin{equation}\label{charact1}
\begin{aligned}
\Delta & =\begin{pmatrix} 1/2 & 1/2 \cr0 & 1/2 \end{pmatrix} ,\quad
\eta _{1}=\begin{pmatrix} 1/2&0\cr0&0 \end{pmatrix}, \quad
\eta _{2}=\begin{pmatrix} 1/2&0\cr1/2&0 \end{pmatrix}, \\
\eta _{3} & =\begin{pmatrix} 0&1/2\cr1/2&0\end{pmatrix}, \quad
\eta _{4}=\begin{pmatrix} 0&1/2\cr1/2&1/2\end{pmatrix}, \quad
\eta _{5}=\begin{pmatrix} 0&0\cr1/2&1/2\end{pmatrix},
\end{aligned}
\end{equation}
and, by convention, $\eta_6$ is the zero theta-characteristic. Note also the property
\begin{equation} \label{chars}
 \eta_1 + \eta_3+ \eta_5 = \eta_2 + \eta_4 = \Delta \quad {\rm mod} \; {\mathbb Z}^{2g}/{\mathbb Z}^{2g} .
\end{equation}

The six functions $\theta\! \left[ \Delta+ \eta_i \right ] \! (z)$, $i=1,\dots,6$ are odd, that is,
$\theta\! \left[ \Delta+ \eta_i \right ] \! (0)=0$,  whereas the other 10 functions $\theta\! \left[ \Delta+ \eta_{ij} \right ] \! (z)$, $i,j\ne 6$
are even. In the case $g=2$ no one of the latter functions vanishes at zero.

\paragraph{The root functions.}
To obtain theta-functions solution for many problems linearized on Jacobians of  hyperelliptic curves,
one can apply some remarkable relations between roots of certain functions on symmetric products of
such curves and quotients of theta-functions with half-integer characteristics, which are historically referred to as {\it root functions}.
For the case of odd order hyperelliptic curves such functions were obtained by Weierstrass and Rosenheim \cite{Weier2, Konigs}, see also \cite{Baker1,BE}.

For our purposes it is sufficient to quote only several root functions for the particular case $g=2$ and the even-order hyperelliptic curve
$$
\Gamma =\{\mu^{2}=R(\lambda )\}, \quad R(\lambda )=(\lambda -E_{1})\cdots(\lambda -E_{6}) .
$$
Let us introduce the polinomial $U(\lambda,s)=(s-\lambda_{1}) (s-\lambda_{2})$.

\begin{proposition} \label{Wurzel}
Under the Abel--Jacobi mapping (\ref{1.16}) with $g=2$ and the basepoint $P_{0}=E_{6}$ the following relations hold
\begin{gather}
\label{root1}
U(\lambda ,E_{i})\equiv (\lambda _{1}-E_{i})(\lambda _{2}-E_{i})
= {\kappa}_i \frac{\theta^2 [\Delta+\eta _{i}](z)}{ \theta [\Delta](z-q/2) \, \theta [\Delta](z+q/2) } ,\\
q=\int^{\infty_+}_{\infty_-} \bar{\omega}=2\int^{\infty _+}_{E_{6}} \bar{\omega } ,
\qquad {\kappa }_{i}=\textup{const} ,\qquad  i=1,\ldots ,6 , \nonumber \\
\frac{1} {\lambda _{1}-\lambda _{2} }
\left( \frac{ \sqrt{R (\lambda_1)} }{ ( E_{i}-\lambda_{1}) ( E_{j}-\lambda_{1}) (E_{s}-\lambda_{1}) }
- \frac{ \sqrt{R (\lambda_2)}  }{ ( E_{i}-\lambda_{2}) (E_{j}-\lambda_{2}) (E_{s}-\lambda_{2}) } \right) \nonumber \\
\qquad \qquad \qquad \qquad = \kappa_{ijs}\, \frac{\theta [\Delta+\eta_{i}+\eta_j+\eta_s] (z)\,  \theta [\Delta](z-q/2) \, \theta [\Delta](z+q/2)  }
{ \theta [\Delta+\eta_i] (z)\, \theta [\Delta+\eta_j] (z)\, \theta [\Delta+\eta_s] (z) } , \label{root20} \\
\frac{ \sqrt{U (\lambda, E_{i} ) }\sqrt{U(\lambda, E_{j})} } {\lambda _{1}-\lambda _{2} }
\left( \frac{ \sqrt{R (\lambda_1)} }{ ( E_{i}-\lambda_{1}) ( E_{j}-\lambda_{1}) (E_{s}-\lambda_{1}) }
- \frac{ \sqrt{R (\lambda_2)}  }{ ( E_{i}-\lambda_{2}) (E_{j}-\lambda_{2}) (E_{s}-\lambda_{2}) } \right) \nonumber \\
\qquad \qquad \qquad \qquad = \kappa_{ijs}'\, \frac{\theta [\Delta+\eta_{i}+\eta_j+\eta_s] (z)}{\theta [\Delta+\eta_s] (z)} ,\label{root2} \\
\kappa_{ijs}, \kappa_{ijs}' =\textup{const} ,\quad
  i,j,s =1,\dots,6 , \quad i\ne j\ne s \ne i, \nonumber
\end{gather}
where, as above, $\eta_6$ is the zero theta-characteristic and $\infty_+,\infty_-$ are the infinite points of the compactified curve $\Gamma$.
The constant factors ${\kappa}_{i}, \kappa_{ijs}, \kappa_{ijs}'$ depend on the moduli of $\Gamma$ only.
\end{proposition}

Note that various expressions of symmetric functions of the $\lambda, \mu$-coordinates on an
even hyperelliptic curve were obtained in \cite{EEH03} on the basis of
the Klein--Weierstrass realization of Abelian functions outlined in \cite{Baker1} and \cite{BEL97}.
\medskip

\noindent{\it Sketch of proof of Proposition} \ref{Wurzel}. The left and right hand sides of \eqref{root1} are meromorphic
functions on Jac$(\Gamma )$, which have the same zeros and poles with the same
multiplicity. This implies that their quotient is an analytic function
on a compact complex manifold without poles and therefore a constant.

The root functions \eqref{root20}, \eqref{root2} can be deduced from the corresponding root functions for the case of odd-order hyperelliptic curve,
by making a fractionally-linear transformation of $\lambda$ that sends the Weierstrass point $E_{2g+2}$ on $\Gamma$ to infinity. $\square$
\medskip

The constants ${\kappa}_{i}, \kappa_{ijs}, \kappa_{ijs}'$ can be calculated explicitly in terms of the coordinates $E_1,\dots, E_6$ and theta-constants
by equating $\lambda_1, \lambda_2$ to certain $E_i$ and the argument $z$ to the corresponding half-period in Jac$(\Gamma)$ (see, e.g., \cite{Baker1}).

\paragraph{Explicit solution.}
Now we are able to write explicit solution for the Steklov--Lyapunov systems by comparing the root functions  (\ref{root1}), (\ref{root2})
with the K\"otter expression \eqref{k8.8}. 

Namely, let $\Gamma =\left\{ \mu ^{2}=\Phi \left( \lambda \right) \varphi \left(\lambda \right) \right\}$
where the polynomials $\phi$ and $\varphi$ are defined in (\ref{fipsi}) and identify (without ordering) the sets
$$
\{ E_{1},E_{2},E_{3},E_{4},E_{5},E_{6}\}= \{ b_{1},b_{2},b_{3},c_{1},c_{2},c_{3}\}.
$$
By $\eta_{b_{i}},\eta_{c_{\alpha}}$ we denote the half-integer characteristics corresponding to the branch points $(b_{i},0), (c_{\alpha },0)$
respectively, according to formula \eqref{eta}.

\begin{theorem}\label{solution}
For fixed constants of motion $c_{1}, c_{2}, c_{3}$ the variables $z, p$
can be expressed in terms of theta-functions of the curve $\Gamma$ in a such a way that for any $s\in {\mathbb C}$
\begin{equation}
z_{i}+sp_{i}=\frac{\sum_{\alpha =1}^{3}k_{i \alpha }\left( s-c_{\alpha }\right) \theta \left[
\Delta +\eta _{c_{\beta }}+\eta _{c_{\gamma }}+\eta _{b_{i}}\right] \left(
z\right) }{\sum_{\alpha =1}^{3} k_{0 \alpha}\, \theta [ \Delta +\eta_{c_\alpha} ] ( z) }, \qquad (\alpha,\beta,\gamma)=(1,2,3), \label{theta_sol}
\end{equation}
where $k_{i \alpha},k_{0 \alpha}$ are certain constants depending on the moduli of $\Gamma$ only,
and the components of the argument $z$ are linear functions of $t$:
\begin{equation} \label{zt}
 z_{1}= C_{11} \delta_1 t+ C_{12}\delta_2 t + z_{10} , \quad  z_{2}= C_{21} \delta_1 t+ C_{22}\delta_2 t + z_{10}, \quad
z_{10}, z_{20}= \textup{const}, \quad C= A^{-1}
\end{equation}
$A$ being is the matrix of $\mathfrak{a}$-periods of the differentials $d\lambda/\mu, \lambda\,d\lambda/\mu$
on $\Gamma$.
\end{theorem}

Thus, we have recovered the theta-function solution of the systems obtained by K\"otter in \cite{Kot2}. The proof
is given in the end of the section.

\paragraph{Remark 2.} In view of the definition of theta-function with characteristics,
under the argument shift $z \to z-{\cal K}$ the special characteristic $\Delta$ is killed and the solutions \eqref{theta_sol} are simplified to
\begin{equation}
z_{i}+sp_{i}=\frac{\sum_{\alpha =1}^{3} \bar k_{i \alpha} ( s-c_{\alpha })
\theta [\eta _{c_{\beta }}+\eta _{c_{\gamma }}+\eta _{b_{i}}] (z) }{ \sum_{\alpha =1}^{3} \bar k_{0 \alpha }\, \theta [\eta_{c_\alpha} ] ( z) },
\qquad (\alpha,\beta,\gamma)=(1,2,3), \label{theta_mod}
\end{equation}
where the constants $\bar k_{i \alpha},\bar k_{0 \alpha}$ coincide with $k_{i \alpha},k_{0 \alpha}$ in \eqref{theta_sol}
up to multiplication by a quartic root of unity.
In each concrete case of position of $b_i, c_\alpha$,
one can also simplify the sums of characteristics in the numerator of \eqref{theta_mod} by using the relations \eqref{chars}.

\paragraph{Remark 3.} In view of the quasi-periodic property \eqref{1.5},
when the complex argument $z$ changes by a period vector in Jac$(\Gamma)$, the theta-functions in \eqref{theta_sol}, \eqref{theta_mod} are multiplied by
generally different factors. Hence, the variables $z_i, p_i$ cannot be single valued on the Jacobian variety
$\Gamma$, and a simple accounting shows that they are meromorphic on
$\widetilde\Jac  (\Gamma)$, the 16-fold unramified covering of it, obtained by doubling all the four period
vectors in Jac$(\Gamma)$. This implies that $\widetilde\Jac  (\Gamma)$ is also
a principally polarized Abelian variety isomorphic to Jac$(\Gamma)$. As follows from the structure of \eqref{theta_sol},
all $z_i, p_i$ have a common set of simple poles (the pole divisor), which we denote ${\cal D}\subset \widetilde\Jac  (\Gamma)$.

The degree of the covering $\widetilde\Jac  (\Gamma) \to \Jac  (\Gamma)$ can also be found in another way: According to Remark 1,  each generic pair
$(\lambda_{1},\lambda_{2})$ corresponds to 64 different points $(z,p)$ on the invariant manifold $\widetilde\Jac  (\Gamma)$.
On the other hand, the same pair gives rise to 4 different
points in Jac$(\Gamma)$ defined by the divisors $\{(\lambda_1, \pm \sqrt{R_6(\lambda_1)}\, ) , (\lambda_2, \pm \sqrt{R_6(\lambda_2)}\,) \,\}$.
Hence a generic point
of Jac$(\Gamma)$ corresponds to 64/4=16 points in $\widetilde\Jac  (\Gamma)$.
\medskip

\noindent{\it Proof of Theorem} \ref{solution}.
The summands in the numerator of the K\"otter solution \eqref{k8.8}, when divided by $\lambda_1-\lambda_2$, can be written as
\begin{gather*}
\frac{s-c_\alpha } { ( c_{\alpha}-c_{\beta} )( c_{\alpha}-c_{\gamma}) }
\frac{ \sqrt{-( \lambda _{1}-c_\alpha) ( \lambda _{2}-c_\alpha )}  }{\lambda_1 -\lambda_2}
 \cdot \biggl( \frac{\sqrt{\Phi (\lambda_1)\psi (\lambda_2)}} {%
(\lambda_1-b_i)(\lambda_2-c_\alpha)} -\frac{\sqrt{\Phi (\lambda_2)\psi
(\lambda_1)}} {(\lambda_2-b_i)(\lambda_1-c_\alpha)} \biggr) \\
\qquad \qquad \qquad = \frac{s-c_\alpha } {( c_{\alpha}-c_{\beta})( c_{\alpha}-c_{\gamma})  }
\frac{ \sqrt{-( \lambda _{1}-c_\beta )(\lambda _{2}-c_\beta )}
\sqrt{-( \lambda_{1}-c_{\gamma})(\lambda_{2}-c_{\gamma} )} }{\lambda_1 -\lambda_2} \\
\quad \qquad \qquad \qquad \qquad \times \left( \frac{\mu_1}{(\lambda_1 -b_i) (\lambda_1-c_\beta) (\lambda_1-c_\gamma)}-
\frac{\mu_2}{(\lambda_2 -b_i) (\lambda_2-c_\beta) (\lambda_2-c_\gamma) } \right) , \\
\mu_1 =\sqrt{\Phi(\lambda_1) \psi(\lambda_1) } , \quad \mu_2 =\sqrt{\Phi(\lambda_2) \psi(\lambda_2) }\, .
\end{gather*}
The right hand sides have the form of the root function \eqref{root2}. Hence, up to a constant factor, they are equal to
$$
(s-c_\alpha) \frac{\theta \left[ \Delta +\eta _{c_\beta}+\eta _{c_\gamma }+ \eta _{b_{i}} \right] (z) }{\theta[\Delta+\eta _{b_{i}}  ](z)}.
$$
Next, in view of (\ref{root1}), we obtain
\begin{align}
x_i & = {\varkappa}_i \frac{\theta [\Delta+\eta_{b_i}](z)}{\sqrt{\theta [\Delta](z-q/2) \, \theta [\Delta](z+q/2)}}\, , \notag \\
 \sqrt{-( \lambda _{1}-c_\alpha) ( \lambda _{2}-c_\alpha )}
& = {\varkappa}_\alpha \frac{\theta [\Delta+\eta_{c_\alpha}](z)}{\sqrt{\theta [\Delta](z-q/2) \, \theta [\Delta](z+q/2)}}\, , \label{*} \\
& \varkappa_i, \varkappa_\alpha = \mbox{const} , \notag \\
\sum_{\alpha=1}^3 \frac{\sqrt{-(\lambda_1-c_\alpha)(\lambda_2-c_\alpha) } } {(c_\alpha -c_\beta)(c_\alpha-c_\gamma)}
& = \sum_{\alpha=1}^3 \dfrac{k_{0 \alpha }\theta [ \Delta +\eta _{c_{\alpha }}]
( z) }{ \sqrt{ \theta [ \Delta ] ( z-q/2) \theta [ \Delta ] (z+q/2) } } . \label{***}
\end{align}
Combining the above expressions, we rewrite the right hand side of (\ref{k8.8}) in the form
\begin{equation*}
\sqrt{c_0}\, \dfrac{\dfrac{ \theta [ \Delta+\eta _{b_{i}} ]( z)
}{ \sqrt{ \theta \left[ \Delta \right] ( z-q/2 ) \theta [ \Delta ]( z+q/2 ) } }
\sum_{\alpha=1}^3 \dfrac{k_{i \alpha}\left( s-c_{\alpha }\right) \theta \left[
\Delta +\eta _{c_\beta}+\eta _{c_\gamma}+\eta _{b_{i}}\right] \left(
z\right) }{\theta \left[ \Delta +\eta _{b_{i}}\right] \left( z\right) }}{
\sum_{\alpha=1}^3 \dfrac{k_{0 \alpha }\theta [ \Delta +\eta _{c_{\alpha }}]
( z) }{ \sqrt{ \theta [ \Delta ] ( z-q/2) \theta [ \Delta ] (z+q/2) } }} \, ,
\end{equation*}
which, after simplifications, gives \eqref{theta_sol}.

Expressions (\ref{zt}) follow from the relation $(\bar\omega_1, \bar\omega_2)^T=C (d\lambda/\mu, \lambda\,d\lambda/\mu)$,
where, as above, $\bar\omega_j$ are the normalized holomorphic differentials on $\Gamma$, which implies
$(\bar\omega_1, \bar\omega_2)^T=C (u_1, u_2)^T$, where $u_1, u_2$ are the right hand sides of the quadratures
\eqref{u_s}. $\square$

\section{The divisor of poles and the alternative form of the theta-function solution.}
The nice form of the K\"otter solution \eqref{theta_sol} itself tells us a little about the structure of zeros and poles of $z_i, p_i$ on the
2-dimensional Abelian variety $\widetilde\Jac  (\Gamma)$. It is possible however to give a quite detailed description of the set of common poles of these variables, called the divisor of poles $\cal D$.
Obviously,
${\cal D}=\{ \sum_{\alpha=1}^3 k_{0 \alpha } \theta [ \Delta +\eta _{c_{\alpha }}]( z)=0 \}\subset\widetilde\Jac  (\Gamma)$.
\medskip

Namely, for each $\alpha=1,2,3$, the zeros of $\theta [\Delta+\eta_{c_\alpha}]( z)$ in Jac$(\Gamma)$ form a translate $\Theta_\alpha$ of the theta-divisor $\Theta$ by
the half-period $2 \pi \jmath \eta''_{c_\alpha} + 2 B\eta'_{c_\alpha}$.
Each translate passes via six half-periods, and $\Theta_1, \Theta_2, \Theta_3$ have a unique common intersection in the origin (neutral point)
${\cal O}\in \Jac (\Gamma)$.
This is depicted in Fig. 3 (a), where $\Theta_\alpha$ are shown as circles and the
half-periods in  Jac$(\Gamma)$ as black dots.  Hence, at $z={\cal O}$ the denominator of \eqref{theta_sol} vanishes. Then, under the covering
$\pi\, : \,\widetilde\Jac  (\Gamma) \to \Jac (\Gamma)$,
the preimage of $\cal O$ consists of all the 16 half-periods in $\widetilde\Jac  (\Gamma)$, which therefore belong to the divisor $\cal D$.

Note that translations in Jac$(\Gamma)$ by a complete period ${\cal V}$ correspond to translation in $\widetilde\Jac  (\Gamma)$ by the half-period ${\cal V}/2$.

Now assume, as above, that $b_1<b_2<b_3$ and that $(b_3,0)=E_6\in \Gamma$ is the basepoint of the Abel map
(\ref{1.16}) with $g=2$.
A further information about $\cal D$ is given by 

\begin{proposition} The divisor ${\cal D}\subset \widetilde\Jac  (\Gamma)$ is invariant under translations by the half-periods generated by
\begin{equation} \label{V12}
{\cal V}_1/2 = 2 \pi \jmath \eta''_{b_1} + 2 B\eta'_{b_1}, \quad {\cal V}_2/2= 2 \pi \jmath \eta''_{b_2} + 2 B\eta'_{b_2}, \qquad
\begin{pmatrix} \eta_{b_i}' \\ \eta''_{b_i} \end{pmatrix}= \eta_{b_i} .
\end{equation}
\end{proposition}

\noindent{\it Proof.} Choose a generic point $q\in {\cal D}$ and let $z^*$ be its projection onto Jac$(\Gamma)$, which gives
$$
f(z^*)=\sum_{\alpha =1}^{3} k_{0 \alpha}\, \theta [\Delta+ \eta_{c_\alpha} ] ( z^*)=0 . 
$$
In view of the quasi-periodic property \eqref{1.5} and the half-integer characteristics \eqref{charact1}, under the translations
$z^* \to z^*+ M {\cal V}_1 + N {\cal V}_2$, $M,N \in {\mathbb Z}$ all the functions $\theta [\Delta+\eta_{c_\alpha} ] ( z^*)$ are multiplied by the same factor
and therefore $ f( z^*+ M {\cal V}_1 + N {\cal V}_2 )=0$. Hence the points $z^*/2+ M {\cal V}_1/2 + N {\cal V}_2/2$ in $\widetilde\Jac  (\Gamma)$ also belong to $\cal D$.

One can also show that this does not hold for the translations by the other half-periods. $\square$

\medskip

\begin{theorem} \label{factor} The denominator of the solution \eqref{theta_sol} admits the factorization
\begin{equation} \label{4_thetas}
 \sum_{\alpha=1}^3 k_{0 \alpha }\theta [ \Delta +\eta _{c_{\alpha }}]( z)=\exp(\chi z+\zeta)\cdot \theta[ \Delta](z/2)\,
\theta [\Delta+\eta_{b_1} ] (z/2)\, \theta [\Delta+\eta_{b_2}](z/2)\, \theta [\Delta+\eta_{b_1}+\eta_{b_2}] (z/2)
\end{equation}
with certain constants $\chi, \zeta$.
\end{theorem}

The proof of the theorem is based on the fourth Riemann identity (see, e.g., \cite{Baker1, Dub_NE}) and
the theta-formulas of Frobenius and Thomae (see, e.g., \cite{Thom, Mum_Theta}). Technically, it is quite tedious and
for this reason we move it into Appendix.
\medskip

Now note that each of the 4 sets
\begin{gather*} %
{\cal D}_0= \{\theta[ \Delta](z/2\, |B)=0  \}, \quad {\cal D}_1= \{\theta [\Delta + \eta_{b_1} ] (z/2\, |B)=0 \}, \\
{\cal D}_2 = \{\theta [\Delta + \eta_{b_2} ] (z/2\, |B)=0 \}, \quad {\cal D}_3= \{ \theta [\Delta+\eta_{b_1}+\eta_{b_2}] (z/2\,|B)=0 \} 
\end{gather*}
describe a translate of the theta-divisor, the genus 2 curve $\Gamma$ embedded into $\widetilde\Jac  (\Gamma)$. Then, Theorem \ref{factor} says that the pole divisor $\cal D$
is a union of these translates, which
are obtained from each other by shifts by the half-periods ${\cal V}_1/2, {\cal V}_2/2$, and ${\cal V}_3/2= -{\cal V}_1/2- {\cal V}_2/2$.
The union passes through all the 16 half-periods in $\widetilde\Jac  (\Gamma)$. The action of the translations by
${\cal V}_1/2, {\cal V}_2/2, {\cal V}_3/2$ in $\widetilde\Jac  (\Gamma)$ on the components $({\cal D}_0, {\cal D}_1, {\cal D}_2, {\cal D}_3)$ gives respectively
\begin{gather} \label{action}
( {\cal D}_1, {\cal D}_0, {\cal D}_3, {\cal D}_2 ), \quad ({\cal D}_2, {\cal D}_3, {\cal D}_0, {\cal D}_1 ), \quad ({\cal D}_3, {\cal D}_2, {\cal D}_1, {\cal D}_0).
\end{gather}

All these properties are in complete correspondence with our previous observations about the divisor $\cal D$.

Also, as was shown in \cite{AvM1} by applying the Kovalevskaya--Painlev\'e analysis, the pole divisor with the same structure appears
in the integrable flow on the algebra $so(4)$ with the diagonal metric II, already mentioned in Introduction.
This result of \cite{AvM1} about $\cal D$ equally holds for the Steklov--Lyapunov systems
due to a linear isomorphism between them and the integrable flow on $so(4)$.

The intersection pattern for $\cal D$  is shown
in Fig. 3 (b), which we borrowed from \cite{AvM1}. Here the circles represent
the translates ${\cal D}_j$ and the 16 black dots depict the half-periods.
Under the projection $\pi$ all the above half-periods are mapped onto ${\cal O} \in \Jac  (\Gamma)$.

\begin{figure}[h,t] 
\begin{center}
\subfigure[]{\includegraphics[height=0.3\textwidth]{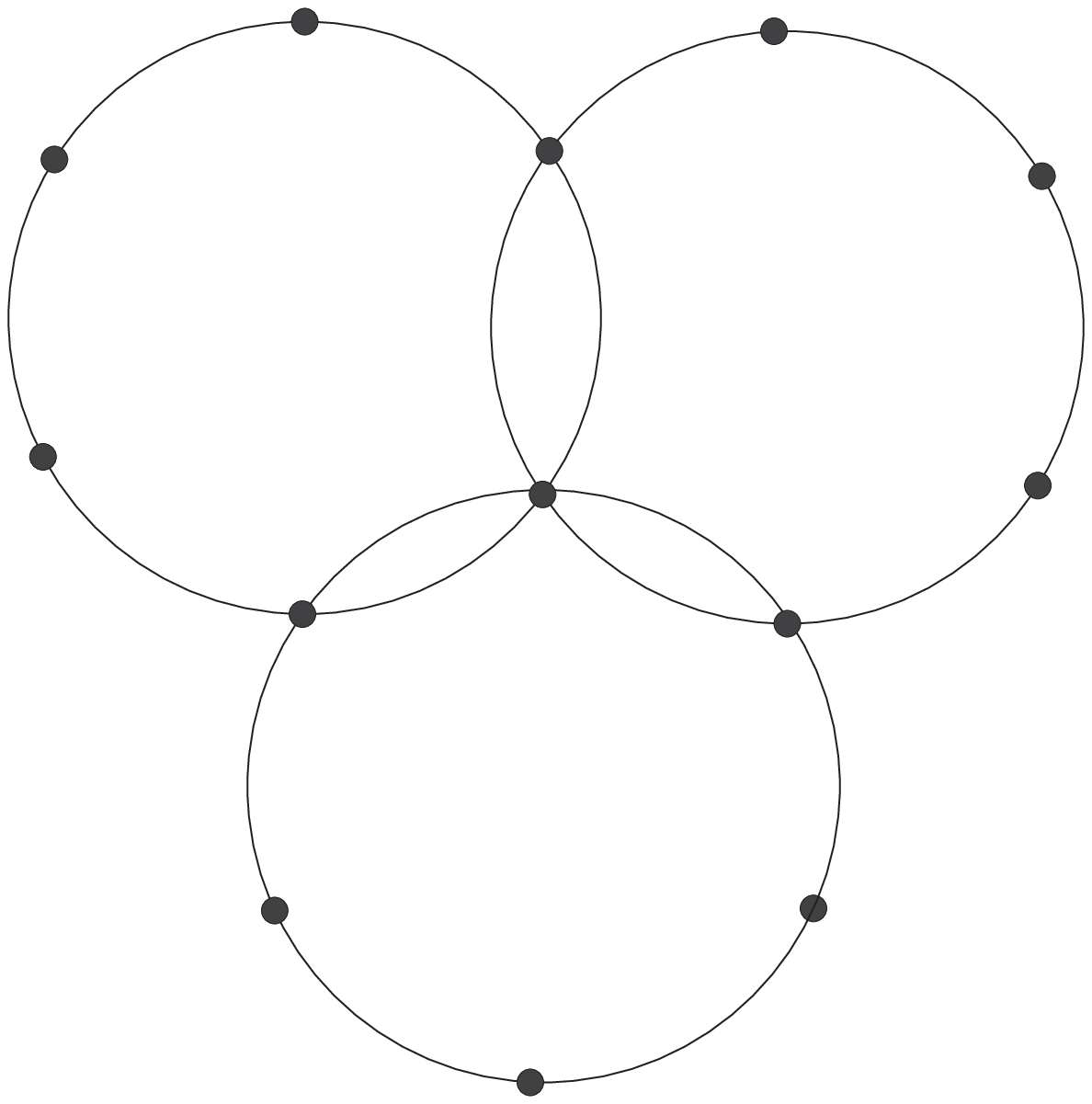}} \quad
\subfigure[]{\includegraphics[height=0.3\textwidth]{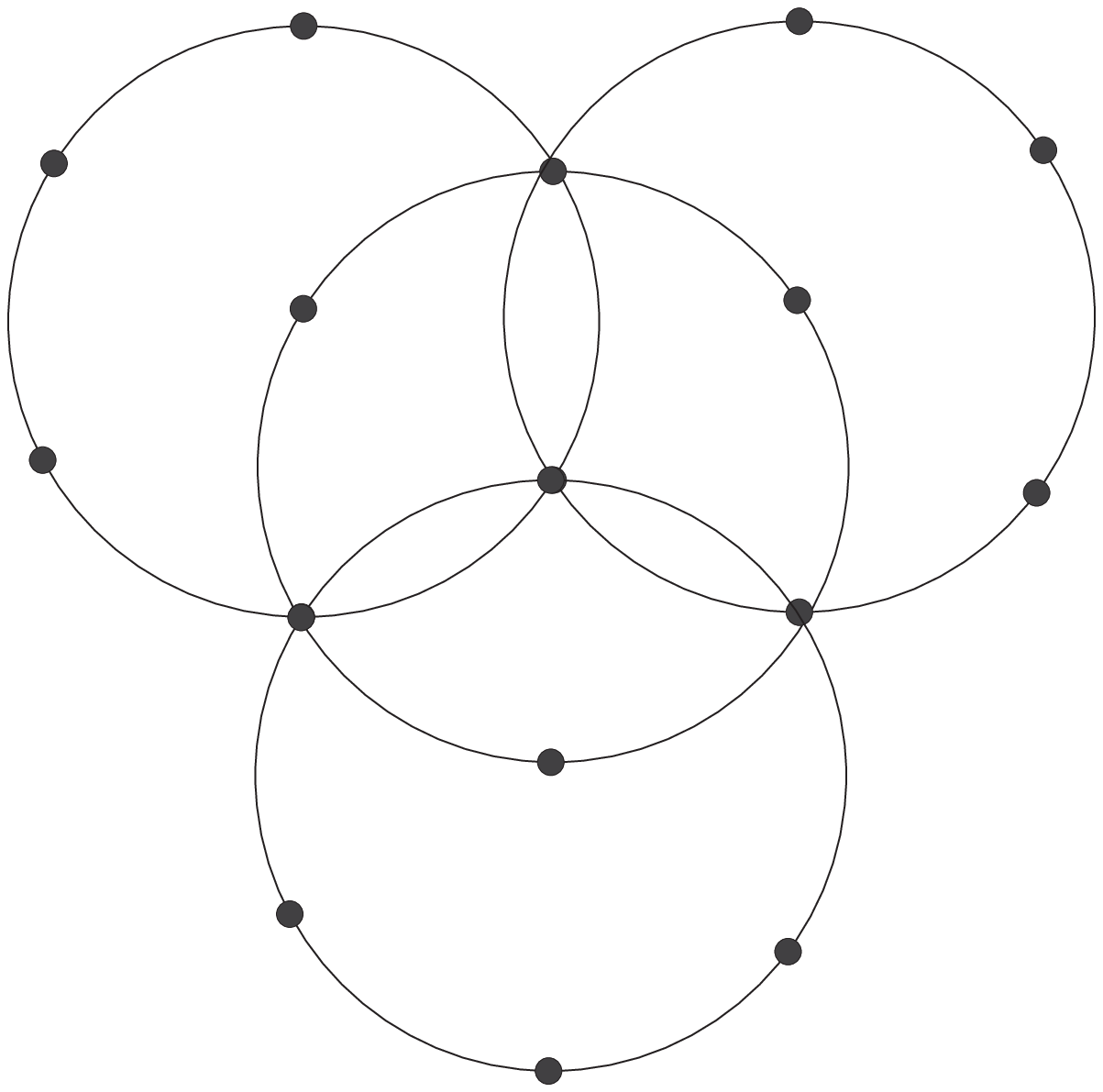}}
\end{center}
\caption{\footnotesize (a) Configuration of the translates $\Theta_\alpha$ in Jac$(\Gamma)$. (b) The 4 translates of $\Gamma$ in $\widetilde\Jac  (\Gamma)$ forming
the pole divisor $\cal D$.}
\end{figure}

\paragraph{Solutions for the variables $v_k$.} Let us choose the origin of $\widetilde\Jac  (\Gamma)$ at one of the four triple intersections of ${\cal D}_j$
and denote for brevity the four theta-functions in \eqref{4_thetas} as
$\theta_0(z/2), \theta_1(z/2), \theta_2(z/2), \theta_3(z/2)$.

Now we show that theta-function solutions for the new phase variables $v_1,\dots, v_6$ introduced in \eqref{vis} have a rather specific and compact form.
Namely, as follows from expressions \eqref{theta_sol} and \eqref{vis}, the functions $v_1+v_2$ and $v_1-v_2$ may have
only {\it simple} poles {\it at most} along the components of the divisor ${\cal D}$. On the other hand, the form of the integrals (\ref{ints_v})
imply the following remarkable property: the poles (the zeros) of $v_1+v_2$ are the zeros (resp. the poles) of $v_1-v_2$.
Since both functions are meromorphic on $\widetilde\Jac  (\Gamma)$, none of then can have simple poles along only one component ${\cal D}_{j}$.
This necessarily implies that $v_1+v_2$ has poles along two certain components
$ {\cal D}_{j_1}, {\cal D}_{j_2}$ and zeros along the other two components $ {\cal D}_{j_3}, {\cal D}_{j_4}$, and vice versa for $v_1-v_2$.

The same observations hold for the pairs $(v_3+v_4$, $v_3-v_4)$ and $(v_5+v_6$, $v_5-v_6)$. Note also that functions from different pairs cannot have
the same poles, since in that case they would also have the same zeros and their quotient would be constant, which is not true.

Now let us fix the origin of $\widetilde\Jac  (\Gamma)$ at one specific triple intersection of ${\cal D}_j$ such that the 3 functions $v_1+v_2, v_3+v_4, v_5+v_6$ have a common
pole along the component ${\cal D}_0$.  In this case the following proposition holds.

\begin{proposition} The theta-function solutions for the phase variables $v_k$ have the form
\begin{gather}
v_1+v_2 = \chi_1\, \frac{ \theta_1 (z/2)\, \theta_2 (z/2) }{\theta_0 (z/2) \, \theta_3 (z/2)},  \quad
v_1-v_2 = \chi_2\, \frac{\theta_0 (z/2) \, \theta_3 (z/2)} { \theta_1 (z/2)\, \theta_2 (z/2) } , \notag \\
v_3+v_4 = \chi_3\, \frac{ \theta_2 (z/2)\, \theta_3 (z/2) }{\theta_0 (z/2) \, \theta_1 (z/2)},  \quad
v_3-v_4 = \chi_4\, \frac{ \theta_0 (z/2) \, \theta_{1}(z/2)} { \theta_2 (z/2)\, \theta_3 (z/2) } , \label{dsols} \\
v_5+v_6 = \chi_5\, \frac{ \theta_1 (z/2)\, \theta_3 (z/2) }{\theta_0 (z/2) \, \theta_2 (z/2)},  \quad
v_5-v_6 = \chi_6\, \frac{\theta_0 (z/2) \, \theta_2 (z/2)} {\theta_1 (z/2)\, \theta_3 (z/2) } , \notag \\
\chi_1, \chi_3, \chi_5 =\textup{const}, \qquad  \chi_2= \frac{\psi(b_1)}{(b_2-b_3)\chi_1},\quad
\chi_4= \frac{\psi(b_2)}{(b_3-b_1)\chi_3}, \quad \chi_6= \frac{\psi(b_3)}{(b_1-b_2)\chi_5} \, , \label{conss}
\end{gather}
where $z=(z_1, z_2)$ depend on $t$ according to \eqref{zt}.
\end{proposition}

\noindent{\it Proof.} First, note that the functions \eqref{dsols} have the same structure of zeros and poles, as
prescribed above.
Next, as follows from the K\"otter formula (\ref{k8.8}) and theta-solutions \eqref{theta_sol}, the translations by the period vectors
${\cal V}_1, {\cal V}_2, {\cal V}_1+ {\cal V}_2$ in Jac$(\Gamma)$  generate the involutions
\begin{align*}
\sigma_1 \; & : \; (z_1, p_1, z_2, p_2, z_3, p_3) \mapsto (z_1, p_1, -z_2, -p_2, z_3, p_3), \\
\sigma_2 \; &: \; (z_1, p_1, z_2, p_2, z_3, p_3) \mapsto (-z_1, -p_1, -z_2, -p_2, z_3, p_3), \\
\sigma_3= \sigma_2\circ \sigma_1 \; &: \; (z_1, p_1, z_2, p_2, z_3, p_3) \mapsto (-z_1, -p_1, z_2, p_2, z_3, p_3) ,
\end{align*}
which, in view of \eqref{vis}, gives rise to the transformations 
\begin{align*}
\sigma_1 \; &: \; v_2 + v_1 \longleftrightarrow v_2 - v_1 , \quad v_4 \pm v_3 \longleftrightarrow v_4 \pm v_3, \quad  v_5 + v_6 \longleftrightarrow v_5 - v_6 , \\
\sigma_2 \; &: \; v_2 + v_1 \longleftrightarrow v_2 - v_1 , \quad v_4 + v_3 \longleftrightarrow v_4 - v_3 , \quad  v_5 \pm v_6 \longleftrightarrow -(v_5 \pm v_6) \\
\sigma_3 \; &: \; v_2 \pm v_1 \longleftrightarrow v_2 \pm v_1 , \quad v_4 + v_3 \longleftrightarrow v_4 - v_3 , \quad  v_6 + v_5 \longleftrightarrow v_6 - v_5 .
\end{align*}
Now observe that the relations \eqref{dsols} are invariant under the action of $\sigma_i$ on the left-hand sides
and the corresponding transformation of $\theta_0(z/2), \dots, \theta_3(z/2)$ under the
action (\ref{action}). Moreover, one can check that the left- and right hand sides of \eqref{dsols}
are multiplied by the same factors under the shift of $z$ by any period vector of Jac$(\Gamma)$. This
proves \eqref{dsols}.

The relations \eqref{conss} between the constants $\chi_i$ follow from the first 3 integrals in \eqref{ints_v}. $\square$
\medskip

The constants $\chi_1, \chi_2, \chi_3$  can be calculated explicitly in terms of $b_i, c_\alpha$ and theta-constants of $\Gamma$.

As follows from the solutions \eqref{dsols}, the product $(v_1+v_2)(v_3+v_4)$ and the other two similar products have double
poles along ${\cal D}_0$ only:
\begin{gather}
(v_1+v_2)(v_3+v_4)= g_2\frac{\theta^2_2 (z/2)}{\theta_0^2(z/2)}, \notag \\
(v_3+v_4)(v_5+v_6)= g_3\frac{\theta^2_3 (z/2)}{\theta^2_0(z/2)}, \quad
(v_1+v_2)(v_5+ v_6)= g_1\frac{\theta^2_1 (z/2)}{\theta_0^2(z/2)} , \notag \\
g_1, g_2, g_3= \mbox{const} . \notag
\end{gather}
Analogs of some of these expressions were obtained in paper \cite{Bueken} in relation with separation of variables for the
integrable system on $so(4)$ with the diagonal metric II. Due to the linear isomorphism between this system and the Steklov--Lyapunov systems,
the separating variables presented in \cite{Bueken} can also be regarded as new separating variables for (\ref{11}), (\ref{22}).

\section{Conclusive Remarks} In given paper we gave a justification of the separation of variables and the theta-function solution of
the Steklov--Lyapunov systems obtained by F. K\"otter \cite{Kot2}. Using the results of
\cite{AvM3, AvM1}, we also presented such a solution for an alternative set of variables, which have a simpler form.

On the other hand, there exist several nontrivial integrable generalizations of the systems: the first of them was discovered by V.
Rubanovsky \cite{Rub2} and describes a motion of a gyrostat
in an ideal fluid under the action of the Archimedes torque, which arises
when the barycenter of the gyrostat does not coincide with its volume
center. In this generalization the Hamiltonian of the Kirchhoff equations, apart form quadratic terms, contains linear
(gyroscopic) terms in $K,p$. Under the change of variables \eqref{substit}, the gyroscopic generalizations of the
systems (\ref{11}), (\ref{22}) take the form
\begin{equation*}
\dot z=z\times (Bz-g)-Bp\times (Bz-g)\, ,\qquad \dot p=p\times (Bz-g)
\end{equation*}
and, respectively,
\begin{equation*}
\dot z=p\times (Bz-g)\, ,\qquad
\dot p=p\times (z-Bp)\,  ,
\end{equation*}
where $g=(g_{1},g_{2},g_{3})^{T}$ is an arbitrary constant vector
related to the angular momentum of the rotor inside the gyrostat.

Following \cite{n-dim_Steklov}, these systems admit the following generalizations of K\"otter's Lax pair with
an elliptic spectral parameter
\begin{gather*}
\dot L(s)= [\, L(s),A(s)\, ]\, ,          \qquad
L(s),A(s)\in so(3),\quad s\in \mathbb{C}\, ,       \\
L(s)_{\alpha \beta} =\varepsilon_{\alpha \beta \gamma}\Bigl( \sqrt{s-b_\gamma}\, (z_\gamma + sp_\gamma)
+ g_{\gamma} /\sqrt{s-b_\gamma}\Bigr)\, , \\
A(s)_{\alpha \beta} =\varepsilon_{\alpha \beta \gamma} \frac{1}{s}
\sqrt{(s-b_\alpha )(s-b_\beta)}\, (b_{\gamma}z_{\gamma}-g_{\gamma})\,  , \quad \mbox{resp.} \quad
A(s)_{\alpha \beta}=\varepsilon_{\alpha \beta \gamma} \sqrt{(s-b_\alpha )(s-b_\beta )}\,  p_{\gamma} ,
\end{gather*}
which provides a sufficient set of constants of motion and makes possible to obtain theta-function solutions.
Like in the case of the Steklov--Lyapunov systems, generic invariant manifolds of the Rubanovsky systems are two-dimensional tori,
which can be extended to affine parts of Abelian varieties.
However, as we plan to show in a forthcoming publication, an explicit integration of the latter systems appears to be more complicated,
and the Abelian varieties are not Jacobians of genus 2 hyperelliptic curves, but Prym subvarieties.

The problem of separation of variables for the Rubanovsky systems is still unsolved.

\section*{Acknowledgements}
The first author (Yu.F.) acknowledges the
support of grant MTM 2006-14603 of the Spanish Ministry of Science and Technology.

\section*{Appendix. {\it Proof of Theorem} \ref{factor}.}
The proof is based on the fourth Riemann identity (see, e.g., \cite{Baker1, Dub_NE})
\begin{equation} \label{RI}
\theta (y_1) \, \theta(y_2) \, \theta(y_3) \, \theta (y_4)
= \frac 14 \sum \theta\! \left[{ \alpha \atop \beta}\right] \! (w_1)\,  \theta\! \left[{ \alpha \atop \beta}\right]\! (w_2)
\theta\! \left[{ \alpha \atop \beta}\right]\! (w_3) \, \theta\! \left[{ \alpha \atop \beta}\right]\! (w_4),
\end{equation}
where the summation is over all the half-period characteristics $\left[{ \alpha \atop \beta}\right]$ and the arguments $y_j, w_j\in {\mathbb C}^g$
(in our case $g=2$) are related as follows
$$
(w_1\, w_2\, w_3\, w_4) = (y_1\, y_2\, y_3\, y_4)\, T, \quad
T= \frac 12 \begin{pmatrix} 1 & 1 & 1 & 1 \\
 			 1 & 1 & -1 & -1 \\
			1 & -1 & 1 & -1 \\
			1 & -1 & -1 & 1 \end{pmatrix} .
$$

Up to multiplication by a simple exponent of $z$, the theta-product in \eqref{4_thetas}  can be written as
\begin{equation} \label{prod}
\theta(z'/2)\, \theta (z'/2+ {\cal V}_1 ) \, \theta (z'/2+{\cal V}_2 )\, \theta (z'/2+ {\cal V}_1 + {\cal V}_2),
\end{equation}
where $z'=z+2{\cal K}$, i.e., the translation by the complete period in Jac$(\Gamma)$, and ${\cal V}_{1,2}$ are the periods defined by \eqref{V12}.
In view of the identity \eqref{RI}, the product \eqref{prod} gives the following sum of 16 theta-products:
$$
\frac 14 \sum_{ 2(\alpha,\beta)\in ({\mathbb Z}_2)^4} \theta\! \left[{ \alpha \atop \beta}\right] \! \left(z'+ \frac {{\cal V}_1+{\cal V}_2}{2} \right)\,
 \theta\! \left[{ \alpha \atop \beta}\right]\! \left(\frac {{\cal V}_1}{2} \right)\,
\theta\! \left[{ \alpha \atop \beta}\right]\! \left(\frac {{\cal V}_2}{2} \right) \, \theta\! \left[{ \alpha \atop \beta}\right]\! (0) \, .
$$
(Note that in each product the variable $z'$ enters only once.)

Next, in view of the property (\ref{*.*}), this sum can be written as product of an exponent of $z$ and the sum
\begin{gather*}
\frac 14 \sum_{ 2(\alpha,\beta)\in ({\mathbb Z}_2)^4} \theta\! \left[ { \alpha +  \alpha' \atop \beta+\beta' } \right] \! \left( z' \right)\,
 \theta\! \left[{ \alpha +  \alpha' \atop \beta+\beta' }\right]\! \left(-\frac {{\cal V}_1}{2} \right)\,
\theta\! \left[ { \alpha +  \alpha' \atop \beta+\beta' } \right]\! \left(-\frac {{\cal V}_2}{2} \right) \,
\theta\! \left[ { \alpha +  \alpha' \atop \beta+\beta' } \right]\!  \left(-\frac {{\cal V}_1+{\cal V}_2}{2} \right)\, , \\
2\pi \jmath \, \beta'+B \alpha' = ({\cal V}_1+{\cal V}_2)/2 ,
\end{gather*}
which, under the corresponding re-indexation, reads
\begin{gather}
\frac 14 \sum_{ 2(\alpha,\beta)\in ({\mathbb Z}_2)^4} \epsilon_{\alpha,\beta}\,  
\theta\! \left[ { \alpha \atop \beta } \right] \! \left( z' \right)\,
 \theta\! \left[{ \alpha \atop \beta }\right]\! \left(\frac {{\cal V}_1}{2} \right)\,
\theta\! \left[ { \alpha \atop \beta} \right]\! \left(\frac {{\cal V}_2}{2} \right) \,
\theta\! \left[ { \alpha \atop \beta} \right]\!  \left(\frac {{\cal V}_1+{\cal V}_2}{2} \right) \notag \\
= - \frac 14 \sum_{i=1 }^6
\theta\! \left[ \Delta+ \eta_i \right] \! ( z' )\,
 \theta\! \left[ \Delta+ \eta_i \right]\! \left(\frac {{\cal V}_1}{2} \right)\,
\theta\! \left[ \Delta+ \eta_i \right]\! \left(\frac {{\cal V}_2}{2} \right) \,
\theta\! \left[ \Delta+ \eta_i \right]\!  \left(\frac {{\cal V}_1+{\cal V}_2}{2} \right) \notag \\
+ \frac 14 \sum_{1 \le i <j \le 5 }
\theta\! \left[ \Delta+ \eta_{ij} \right] \! ( z' )\,
 \theta\! \left[ \Delta+ \eta_{ij}  \right]\! \left(\frac {{\cal V}_1}{2} \right)\,
\theta\! \left[ \Delta+ \eta_{ij}  \right]\! \left(\frac {{\cal V}_2}{2} \right) \,
\theta\! \left[ \Delta+ \eta_{ij} \right]\!  \left(\frac {{\cal V}_1+{\cal V}_2}{2} \right) ,
\label{sum_16}
\end{gather}
where $\epsilon_{\alpha,\beta} = -1$ if $\theta[\alpha\, \beta](z)$ is odd and $+1$ otherwise, and, as above,
$\eta_{ij}=\eta_j+\eta_j$ mod $ {\mathbb Z}^2 /{\mathbb Z}^2$.

In fact, most of the theta-constants in \eqref{sum_16} are proportional to $\theta[\Delta+ \eta_i](0)$, $i=1,\dots, 6$ and therefore vanish.
Namely, in the first sum in the right hand side of \eqref{sum_16} all the theta-constants are non-zero if and only if $\eta_i$ is different from
$\eta_{b_1}, \eta_{b_2}$, and $\eta_6=0$. In the second sum, if $\eta_i$ or $\eta_j$ coincides with $\eta_{b_1}$ or $\eta_{b_2}$, then either the first or the second
theta-constant is zero. Otherwise, if $\{\eta_i, \eta_j \}\cap  \{\eta_{b_1}, \eta_{b_2} \}= \emptyset$, then, in view of the relations \eqref{chars}, the third
theta-constant is proportional to $\theta[\Delta+ \eta_k](0)$, for a certain $k\in \{1,\dots, 6\}$ and, therefore, equals zero.

Since for the case of genus 2 there are no even theta-functions which vanish for zero value of the argument (see \cite{Baker1}), one concludes that the above sum contains only 3 non-zero theta-products:
\begin{gather*}
 \sum_{\eta_i \ne \eta_{b_1}, \eta_{b_2}, 0} \theta [\Delta+ \eta_i] (z') \, \theta\! \left[ \Delta+ \eta_i \right]\! \left(\frac {{\cal V}_1}{2} \right)\,
\theta\! [ \Delta+ \eta_i ]\! \left(\frac {{\cal V}_2}{2} \right) \,
\theta\! [ \Delta+ \eta_i ]\!  \left(\frac {{\cal V}_1+{\cal V}_2}{2} \right) \\
=\sum_{\alpha=1}^3 \theta [\Delta+ \eta_{c_\alpha} ] (z') \, \theta\! \left[ \Delta+ \eta_{c_\alpha} \right]\! 
\left(\frac {{\cal V}_1}{2} \right)\, \theta\! \left[ \Delta+ \eta_{c_\alpha} \right]\! \left(\frac {{\cal V}_2}{2} \right) \,
\theta\! \left[ \Delta+ \eta_{c_\alpha} \right]\!  \left(\frac {{\cal V}_1+{\cal V}_2}{2} \right) . 
\end{gather*}

Now, assume (for the moment) the following ordering of the Weierstrass points:
\begin{equation}\label{order1}
E_1=b_1, \quad E_2=c_1, \quad E_3=b_2, \quad E_4= c_2, \quad E_5=c_3.
\end{equation}
Then, in view of \eqref{*.*} and the identities \eqref{chars}, the above sum, up to a constant common factor, can be written as
\begin{gather}
\varepsilon_1 \theta [\Delta+ \eta_{c_1}] (z') \, \theta [\eta_{b_1}+\eta_{c_2}] (0) \, \theta [\eta_{b_2}+\eta_{c_2}] (0) \, \theta [\eta_{c_1}+\eta_{c_3}] (0) \,  \notag \\
\qquad +\varepsilon_2 \theta [\Delta+ \eta_{c_2}] (z') \, \theta [\eta_{b_1}+\eta_{c_1}] (0) \,  \theta [\eta_{b_2}+\eta_{c_1}] (0) \, \theta [\eta_{c_2}+\eta_{c_3}] (0) \notag \\
+\varepsilon_3 \theta [\Delta+ \eta_{c_3}] (z') \, \theta [\eta_{b_2}] (0) \, \theta [\eta_{b_1}] (0) \, \theta [0] (0)\, , \label{sum_3}
\end{gather}
where now $\varepsilon_\alpha$ are certain quartic roots of 1.

Now we are going to show that the denominator in the theta-function solution \eqref{theta_sol}
coincides with \eqref{sum_3} up to multiplication by an exponent of $z'$.

Namely, in view of \eqref{***}, the sum
$ \sum_{\alpha=1}^3 k_{0 \alpha }\theta [ \Delta +\eta _{c_{\alpha }}]( z) $ can be written as a product of
$$
\mbox{const}\, \sqrt{ \theta [ \Delta ] ( z-q/2) \theta [ \Delta ] (z+q/2) }
$$
and the expression
$$
{\cal G}= \frac{ \sqrt{-(\lambda_1-c_1)(\lambda_2-c_1)}} {\sqrt{(c_1 -c_2)(c_1-c_2)} } \sqrt{ \frac{c_2-c_3}{c_1-c_2} }+
\frac{\sqrt{-(\lambda_1-c_2)(\lambda_2-c_2)}} {\sqrt{(c_2 -c_1)(c_2-c_3)} } \sqrt{ \frac{c_3-c_1} {c_1-c_2 } }
+ \frac{\sqrt{-(\lambda_1-c_3)(\lambda_2-c_3)}} {\sqrt{(c_3 -c_1)(c_3-c_2)} }
$$
(there is no second radical in the third summand !). Now, make the projective transformation
$\lambda\to \nu= \lambda/(\lambda-E_6)=\lambda/(\lambda-b_3)$, which sends the Weierstrass points
$c_\alpha, b_1, b_2, b_3$ on $\Gamma$ to $\bar c_\alpha, \bar b_1, \bar b_2$, and $\infty$. The two infinnite points over $\lambda=\infty$ are
mapped to 2 points over $\nu=1$. This change leaves
the sum $\cal G$ almost invariant: it becomes the product of $\mbox{const}/\sqrt{ (\nu_1-1)(\nu_2-1) }$ and the sum
$$
\bar {\cal G} = \frac{ \sqrt{-(\nu_1-\bar c_1)(\nu_2-\bar c_1)}} {\sqrt{(\bar c_1 -\bar c_2)(\bar c_1-\bar c_2)} }
\sqrt{ \frac{\bar c_2-\bar c_3}{\bar c_1-\bar c_2} }+
\frac{\sqrt{-(\nu_1-\bar c_2)(\nu_2-\bar c_2)}} {\sqrt{(\bar c_2 -\bar c_1)(\bar c_2-\bar c_3)} } \sqrt{ \frac{\bar c_3-\bar c_1} {\bar c_1-\bar c_2 } }
+ \frac{\sqrt{-(\nu_1-\bar c_3)(\nu_2-\bar c_3)}} {\sqrt{(\bar c_3 -\bar c_1)(\bar c_3-\bar c_2)} }\, .
$$
Under the Abel map (\ref{1.16}), the radicals in $\bar {\cal G}$ can be expressed completely in terms of the theta-functions and theta-constants of $\Gamma$:
Applying the theta-formulae of Frobenius and Thomae for the case when one of the Weierstrass points of the curve lies at infinity
(see, e.g., \cite{Thom, Mum_Theta}) and keeping the ordering \eqref{order1}, we have
\begin{align}
\frac{ \sqrt{ (\nu_1-\bar c_\alpha)(\nu_2-\bar c_\alpha) }} {\sqrt{(\bar c_\alpha -\bar c_\beta)(\bar c_\alpha-\bar c_\gamma)} }
 & = \pm \varrho_1 \frac {\theta [\Delta + \eta_{c_\beta} +  \eta_{\bar c_\gamma} ](0) }{ \theta [\Delta+ \eta_{c_1} +  \eta_{c_2} + \eta_{c_3} ](0) }
  \, \frac { \theta [\Delta + \eta_{c_\alpha} ](z) }{ \theta [\Delta](z) } , \quad (\alpha,\beta,\gamma)=(1,2,3),  \nonumber \\
 \sqrt{ \frac{\bar c_2-\bar c_3}{\bar c_1-\bar c_2} } & = \varrho_2 \frac{ \theta[ \eta_{c_2}+ \eta_{b_2} ](0)\, \theta[ \eta_{c_2}+ \eta_{b_1}](0) }
{\theta [\eta_{c_1} + \eta_{c_2} + \eta_{c_3} + \eta_{b_2}](0)\, \theta[\eta_{c_1} +  \eta_{c_2} + \eta_{c_3} + \eta_{b_1}](0)   } \, , \label{theta_sums} \\
\sqrt{\frac{\bar c_3-\bar c_1} {\bar c_1-\bar c_2 } } & = \varrho_3 \frac{ \theta[\eta_{c_1}+ \eta_{b_1} ](0)\, \theta[ \eta_{c_1}+ \eta_{b_2} ](0) }
{\theta [\eta_{c_1} + \eta_{c_2} + \eta_{c_3} + \eta_{b_2}](0)\, \theta[\eta_{c_1} +  \eta_{c_2} + \eta_{c_3} + \eta_{b_1}](0)   } , \nonumber
\end{align}
where $\eta_{c_\alpha}, \eta_{b_1}, \eta_{b_2}$ are the same as above and $\varrho_i$ are the appropriate quartic roots of 1. Lastly, we have
\begin{equation} \label{root_0}
\sqrt{ (\nu_1-1)(\nu_2-1) } = \mbox{const}\, \frac { \sqrt{ \theta \left[ \Delta \right] ( z-q/2 )\, \theta [ \Delta ]( z+q/2 ) } }{\theta[\Delta](z) } ,
\end{equation}
where $q$ is the same as in \eqref{root1}, or, in terms of the new coordinate $\nu$ on $\Gamma$, $q/2= \int_\infty^{(1,0)}\bar\omega$.

Combining the above expressions, we see that in the quotient ${\cal G}=\bar {\cal G}/\sqrt{ (\nu_1-1)(\nu_2-1) }$ the term $\theta [\Delta](z)$ is canceled
and in the product  ${\cal G} \sqrt{ \theta [ \Delta ] ( z-q/2) \theta [ \Delta ] (z+q/2) }$ the square root \eqref{root_0} is canceled.
Now, simplifying the theta-characteristics in \eqref{theta_sums} by using \eqref{chars} and ignoring common constant factors, we eventually find
\begin{align}
\mbox{const}\, & \sum_{\alpha=1}^3 k_{0 \alpha }\theta [ \Delta +\eta _{c_{\alpha }}]( z) \notag \\
& = \bar \varepsilon_1 \theta [\Delta + \eta_{c_1}](z) \, \theta [ \eta_{c_1} + \eta_{c_3}](0)\, \theta [ \eta_{c_2} + \eta_{b_1}](0)\, \theta [ \eta_{c_2} + \eta_{b_2}](0) \notag \\
& \quad +  \bar\varepsilon_2 \theta [\Delta + \eta_{c_2}](z) \, \theta [ \eta_{c_2} + \eta_{c_3}](0)\, \theta [ \eta_{c_1} + \eta_{b_1}](0)\, \theta [ \eta_{c_1} + \eta_{b_2}](0)
\notag \\
& \quad + \bar\varepsilon_3 \theta [\Delta + \eta_{c_3}](z) \, \theta [ \eta_{b_2}] (0)\, \theta [\eta_{b_1}](0)\, \theta (0),   \label{sum30}
\end{align}
$\bar\varepsilon_i$ also being certain quartic roots of 1.
The latter expression have the same structure as the sum \eqref{sum_3}.
Lastly, note that under the shift of $z$ by an appropriate complete period in Jac$(\Gamma)$ the roots $\bar\varepsilon_i$ can be made proportional to any
combination of roots $\varepsilon_\alpha$ in \eqref{sum_3}. (This corresponds to choosing an appropriate origin in $\widetilde\Jac  (\Gamma)$.)
Hence, we proved the theorem for the chosen ordering \eqref{order1}.

To complete the proof for the other possible orderings of $b_i, c_\alpha$ it remains to modify the theta-characteristics in \eqref{sum_3}, \eqref{sum30}.
$\square$


\begin{thebibliography}{90}

\bibitem{AvM3}  Adler M., van Moerbeke P.: Geodesic flow on $so(4)$ and the intersections
of quadrics. {\it Proc.Natl. Acad. Sci. USA.\/} {\bf 81}, (1984), 4613--4616

\bibitem{AvM1}  Adler M., van Moerbeke P.
The complex geometry of the Kowalevski--Painlev\'e analysis.
{\it Invent. Math.\/} {\bf 97} (1989), 3--51

\bibitem{Baker1} Baker H.F. {\it Abels Teorem and the Allied Theory Incluiding the Theory of
Theta Functions. \/} Cambridge Univ. Press, Cambridge, 1897

\bibitem{BE} Belokolos E.D., Bobenko A.I., Enol'skii V.Z., Its A.R., and   Matveev V.B.
{\it Algebro-Geometric Approach to Nonlinear   Integrable Equations.\/}
 Springer Series in Nonlinear Dynamics.   Springer--Verlag 1994.

\bibitem{Bob} Bobenko A.I. Euler equations in the Lie algebras $e(3)$ and $so(4)$.
Isomorphisms of integrable cases.{\it Funkts. Anal. Prolozh.\/}  {\bf 20}, No.1,\,
64--66 (1986). English transl.:{\it Funct. Anal. Appl.\/}  {\bf 20} (1986), 53--56

\bibitem{Bols_Fed} Bolsinov A.V., Fedorov Yu.N. Multidimensional integrable
generalizations of the Steklov-Liapunov system.{\it Vestnik Mosk. Univ.,
 Ser. I, Math. Mekh.\/} No.6, 53-56 (1992) (Russian)

\bibitem{Bols_Fed_2} Bolsinov A.V., Fedorov Yu.N. Steklov--Lyapunov type systems. {\it Preprint}

\bibitem{BEL97} Buchstaber V.M., Enolskii V.Z., and Leykin D.V.
{K}leinian functions, hyperelliptic {J}acobians and Aplications",
In: Reviews in Mathematics and Mathematical Physics.
Editors: S. P. Novikov and I. M. Krichever,
Vol. {\bf 10:2}. London, Gordon and Breach, 1--125, 1997.

\bibitem{Bueken} Bueken P, Vanhaecke P.
The moduli problem for integrable systems: the example of a geodesic flow on ${\rm SO}(4)$. {\it J. London Math. Soc.} (2) {\bf 62} (2000), no. 2, 357--369

\bibitem{Dub_NE} Dubrovin B.A.: Theta-functions and non-linear equations. {\it Usp.Mat. Nauk\/}
{\bf 36}, No.2 (1981), 11-80. English transl.: {\it Russ. Math. Surveys\/}.
{\bf 36} (1981), 11--92

\bibitem{EEH03} Eilbeck J.C. and Enolskii V.Z.,  and Holden H.
\newblock The hyperelliptic $\zeta$- functions and the integrable
massive {T}hirring system, {\em Proc. London Math. Soc. A}, {\bf
459}, 1581-1610,  2003.

\bibitem{n-dim_Steklov} Fedorov Yu. Integrable systems, Poisson pencils,
and hyperelliptic Lax pairs. {\it Regul. Chaotic Dyn.\/} {\bf 5} (2000), no. 2, 171--180

\bibitem{Grif_Harr}  Griffiths Ph., Harris J. {\it Principles of Algebraic Geometry.\/}
Wiley Interscience, New York 1978\,

\bibitem{Kolos} Kolosoff G.V. Sur le mouvement d'un corp solide dans un liquide
ind\'efini. C.R. Acad. Sci. Paris {\bf 169} (1919), 685--686

\bibitem{Konigs}  K\"onigsberger L.
Zur Transformation der Abelschen Functionen erster Ordnung.
{\it J. Reine Agew. Math.\/} {\bf 64} (1894), 3--42


\bibitem{Kot2} K\"otter F. Die von Steklow und Liapunow entdeckten integralen F\"alle,
der Bewegung eines starren K\"orpers in einer Fl\"ussigkeit.
{\it Sitzungsber., K\"onig. Preuss. Akad. Wiss., Berlin\/} {\bf 6} (1900), 79--87

\bibitem{Lyap2}  Lyapunov A.M.  New integrable case of the equations of motion of a rigid body in a fluid.
{\it Fortschr. Math.\/} {\bf 25} (1897), 1501-1504   (Russian)

\bibitem{Mum_Theta}
 Mumford D. Tata Lectures on Theta II. {\it Progress in Math.\/}{\bf 43}, 1984

\bibitem{Rub2} Rubanovsky V. Integrable cases in the problem of a heavy solid
 moving in a fluid. Dokl. Akad. Nauk SSSR, {\bf 180},  (1968), 556--559 (Russian).
 English transl.: {\it Sov. Phys., Dokl.\/} {\bf 13} (1968),  395--397

\bibitem{Steklov} Steklov V. On the motion of a rigid body in a fluid. Kharkov, 1903 (Russian)

\bibitem{Thom} Thomae A. Beitrag zur Bestimmung von $\theta(0,\dots,0)$ durch die Klassenmoduln
algebraischen Funktionen. {\it J.Reine Angew. Math.} {\bf 71} (1870)

\bibitem{Tsiganov2} Tsiganov A. V. On the Steklov--Lyapunov case of the rigid body motion.
{\it Regul. Chaotic Dyn}. {\bf 9} (2004), no. 2, 77--89

\bibitem{Weier2} Weierstrass K.
{\it Mathematische Werke I}, vol. 1, 1894

\end{thebibliography}
\end{document}